\documentclass[twocolumn,aps,prb,superscriptaddress,showkeys,showpacs]{revtex4}
\usepackage{bm}
\usepackage{graphicx}
\catcode`@=11
\def\matrix#1{\null\,\vcenter{\normalbaselines\m@th
    \ialign{\hfil$##$\hfil&&\quad\hfil$##$\hfil\crcr
      \mathstrut\crcr\noalign{\kern-\baselineskip}
      #1\crcr\mathstrut\crcr\noalign{\kern-\baselineskip}}}\,}
\def\pmatrix#1{\left(\matrix{#1}\right)}
\def\eqalign#1{\null\,\vcenter{\openup\jot\m@th
\ialign{\strut\hfil$\displaystyle{##}$&$\displaystyle{{}##}$\hfil
\crcr#1\crcr}}\,}
\def\Journal #1,#2,#3,#4#5#6#7{#1 {\bf #2}, #3 (#4#5#6#7)}
\def\lsim{\lower -0.3ex \hbox{$<$} \kern -0.75em \lower 0.7ex \hbox{$\sim$}}
\def\gsim{\lower -0.3ex \hbox{$>$} \kern -0.75em \lower 0.7ex \hbox{$\sim$}}
\catcode`@=12
\begin{document}
%
\keywords{carbon nanotubes, flattened carbon nanotubes, effective-mass scheme, electronic states, inter-wall interaction}
\pacs{61.48.De, 71.20.Nr, 81.05.U-, 81.05.ue}
%
\title{Effective-mass theory of collapsed carbon nanotubes}
\bigskip
\author{Takeshi Nakanishi}
\email{t.nakanishi@aist.go.jp}
\affiliation{Nanosystem Research Institute, AIST, 1--1--1 Umezono, Tsukuba 305--8568, Japan}
\author{Tsuneya Ando}
\affiliation{Department of Physics, Tokyo Institute of Technology, 2--12--1 Ookayama, Meguro-ku, Tokyo 152--8551, Japan}
\date{\today}
%
\begin{abstract}
Band structure is theoretically studied in partially flattened carbon nanotubes within an effective-mass scheme.
Effects of inter-wall interactions are shown to be important in non-chiral nanotubes such as zigzag and armchair and can essentially be neglected in chiral nanotubes except in the close vicinity of non-chiral tubes.
In fact, inter-wall interactions significantly modify states depending on relative displacement in the flattened region in non-chiral tubes and can convert semiconducting tubes into metallic and vice versa.
They diminish rapidly when the chiral angle deviates from that of the zigzag or armchair tube, although the decay is slower in the vicinity of armchair tubes.
\end{abstract}
%
\maketitle
%
\section{Introduction} \label{Sec:Introduction}
%
Carbon nanotubes were first found in a form of multi-wall cylinders, each of which consists of a rolled graphene sheet.\cite{Iijima_1991a,Iijima_et_al_1992a}
A single-wall nanotube, fabricated later,\cite{Iijima_and_Ichihashi_1993a,Bethune_et_al_1993a} has a unique electronic property that it changes critically from metallic to semiconducting depending on its tubular circumferential vector.
This characteristic feature was first predicted by means of tight-binding models,\cite{Hamada_et_al_1992a,Mintmire_et_al_1992a,Saito_et_al_1992a,Dresselhaus_et_al_1992a,Dresselhaus_et_al_1992b,Jishi_et_al_1993a,Tanaka_et_al_1992a,Gao_and_Herndon_1992a,Robertson_et_al_1992a,White_et_al_1993a} and was successfully described in an effective-mass approximation.\cite{Ajiki_and_Ando_1993a,Ajiki_and_Ando_1996a,Ando_2005a_and_References}
Experimental\cite{Chopra_et_al_1995b,Bourgeois_and_Bursill_1997a,Benedict_et_al_1998a,Liu_et_al_2002d,Li_et_al_2007a,Zhong_et_al_2012a,Kohno_et_al_2013a,Zhang_et_al_2012a,Martel_et_al_1998b,Hertel_et_al_1998a,Yu_et_al_2001d,Yu_et_al_2001b,Giusca_et_al_2007a,Giusca_et_al_2008a,Choi_et_al_2013a} as well as computational studies\cite{Crespi_et_al_1996a,Crespi_et_al_1998a,Lu_et_al_2003a,Liu_and_Cho_2004a,Tangney_et_al_2005a,Lammert_et_al_2000a,Mehrez_et_al_2005a,Lu_et_al_2005a,Tang_et_al_2005a,Zhang_et_al_2006b,Hasegawa_and_Nishidate_2006a,Xiao_et_al_2007a,Lu_et_al_2011a,Kim_et_al_2001e,Nishidate_and_Hasegawa_2008a,Shtogun_and_Woods_2009a,Hasegawa_and_Nishidate_2009a,Nishidate_and_Hasegawa_2010a,Shklyaev_et_al_2011a,Kou_et_al_2013a} have discovered that large diameter nanotubes have an additional stable flattened structure.
The purpose of this work is to study electronic structure of collapsed carbon nanotubes for arbitrary chirality within the effective-mass approximation.
\par
%
The observation of fully collapsed multi-wall carbon nanotubes was reported in transmission electron microscopy,\cite{Chopra_et_al_1995b,Bourgeois_and_Bursill_1997a,Benedict_et_al_1998a,Liu_et_al_2002d,Li_et_al_2007a,Zhong_et_al_2012a,Kohno_et_al_2013a,Zhang_et_al_2012a} atomic force microscopy,\cite{Zhang_et_al_2012a,Martel_et_al_1998b,Hertel_et_al_1998a,Yu_et_al_2001d} and scanning tunneling microscopy.\cite{Giusca_et_al_2007a,Giusca_et_al_2008a,Yu_et_al_2001b}
Multi-wall nanotubes was shown to exhibit structural deformations in FET devices.\cite{Martel_et_al_1998b}
Recently, high-yield fabrication of high quality collapsed tubes was reported, using solution-phase extraction of inner tubes from large-diameter multi-wall tubes.\cite{Choi_et_al_2013a}
\par
%
Actually, it is shown theoretically by first-principles energy minimization that both flattened and cylindrical nanotubes are stable or meta-stable and the energy of flattened tube is lower than cylindrical tubes with large diameter.\cite{Tang_et_al_2005a,Zhang_et_al_2006b,Hasegawa_and_Nishidate_2006a,Xiao_et_al_2007a,Lu_et_al_2011a}
The cylindrical nanotubes collapse into flattened tubes with a barbell-like cross section under hydrostatic pressure or in the presence of injected charge shown by molecular dynamics simulations.\cite{Liu_and_Cho_2004a,Tangney_et_al_2005a}
Electronic states were studied for collapsed armchair tubes in a tight-binding model\cite{Lammert_et_al_2000a} and for collapsed zigzag tubes by density-functional calculations,\cite{Kim_et_al_2001e,Nishidate_and_Hasegawa_2008a} which demonstrated drastic modification in the energy region close to the Fermi level due to inter-wall interaction.
\par
%
Transport of crossed nanotube junctions results from interacting individual tubes and has been studied both experimentally\cite{Postma_et_al_2000a,Fuhrer_et_al_2000b,Yoneya_et_al_2002a,Gao_et_al_2004a,Znidarsic_et_al_2013a} and theoretically.\cite{Nakanishi_and_Ando_2001a,Maarouf_and_Mele_2011a,Yoon_et_al_2001a,Komnik_and_Egger_2001a}
The conductance is found to depend strongly on the crossing angle with large maxima at commensurate stacking of lattices of two nanotubes.\cite{Nakanishi_and_Ando_2001a,Maarouf_and_Mele_2011a}
A deformation of crossed carbon nanotubes, which may significantly affect the tunneling conductance between nanotubes, has been calculated.\cite{Fuhrer_et_al_2000b,Hertel_et_al_1998a,Yoon_et_al_2001a}
Furthermore, a pseudogap has been predicted to appear for an orientationally ordered crystal of nanotubes due to inter-tube transfer.\cite{Delaney_et_al_1998a,Kwon_et_al_1998a,Maarouf_et_al_2000a,Kwon_and_Tomanek_1998a}
\par
%
Effects of inter-wall interactions in multi-wall nanotubes were also studied.
In general, the lattice structure of each nanotube is incommensurate with that of adjacent walls.\cite{Kociak_et_al_2002a,Zuo_et_al_2003a}
This makes inter-wall electron hopping negligibly small as a result of the cancellation of inter-wall coupling in the absence of disorder.\cite{Yoon_et_al_2002a,Triozon_et_al_2004a,Uryu_and_Ando_2005c,Charlier_et_al_2007a,Uryu_and_Ando_2007e}
In fact, inter-wall hopping integrals vary quasi-periodically from site to site and their average over the distance of the order of the circumference vanishes.
This property was extensively used for theoretical calculations of excitons in double-wall nanotubes.\cite{Tomio_et_al_2012a,Tomio_et_al_2012c}
Further, it is closely related to very weak interlayer interactions in twisted bi- and/or multi-layer graphenes.
\par
%
Experimentally, each layer of some of epitaxially fabricated graphenes having many layers is known to behave almost as a monolayer.\cite{de_Heer_et_al_2007a,Sadowski_et_al_2006a,Sadowski_et_al_2007a,Wu_et_al_2007b,Hass_et_al_2007a,Hass_et_al_2008a,Sprinkle_et_al_2009a,de_Heer_et_al_2010a,de_Heer_et_al_2011a,Orlita_et_al_2008n,Miller_et_al_2009a}
Further, the electronic structure of twisted bilayer graphene with nearly incommensurate lattice structure, both theoretically calculated\cite{Lopes_dos_Santos_et_al_2007a,Latil_et_al_2007a,Hass_et_al_2008a,Shallcross_et_al_2008a,Bistritzer_and_MacDonald_2011a,Bistritzer_and_MacDonald_2011b,Lopes_dos_Santos_et_al_2012a,Moon_and_Koshino_2012a,Moon_and_Koshino_2013a,Lu_and_Fertig_2014a,Mele_2010a,Mele_2011a} and experimentally observed,\cite{Sprinkle_et_al_2009a,Ni_et_al_2008c,Schmidt_et_al_2010a,Luican_et_al_2011a,Brihuega_et_al_2012a,Li_et_al_2010a} shows a linear band dispersion near the charge neutrality point, suggesting weak interlayer interaction.
On the contrary, the interlayer interaction drastically changes electronic states in displaced bilayer graphene having a commensurate lattice structure.\cite{Son_et_al_2011a}
The end of bilayer graphene can be closed and was observed experimentally after thermal treatment.\cite{Liu_et_al_2009c}
Geometry and electronic structure of bilayer graphene with a closed edge were studied by a density functional calculation.\cite{Feng_et_al_2009a}
\par
%
\begin{figure*}
\begin{center}
\begin{minipage}[t]{7.50cm}
\begin{center}
\includegraphics[width=7.50cm]{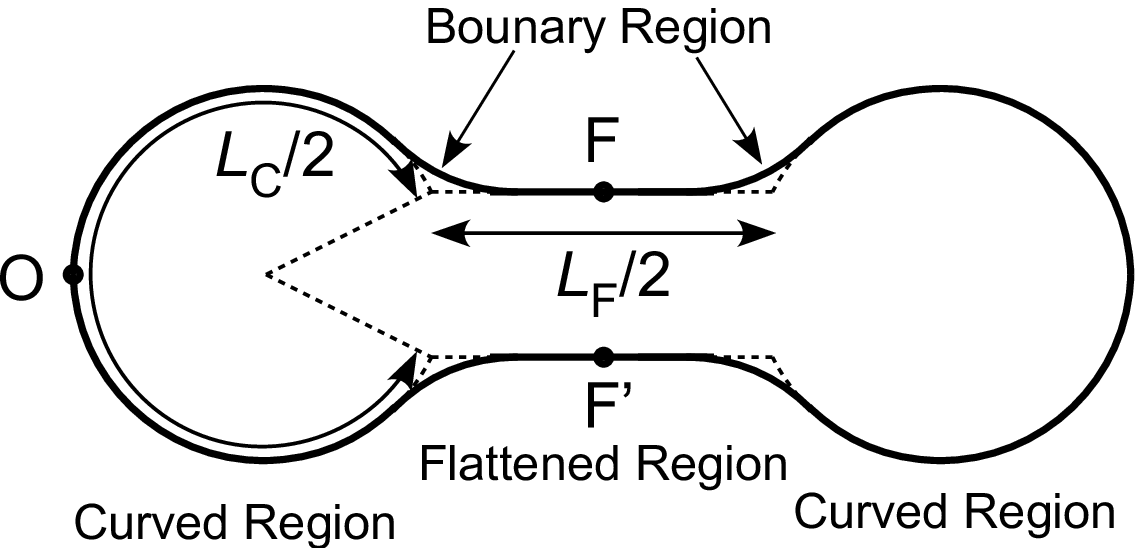}
(a)
\end{center}
\end{minipage}
\begin{minipage}[t]{9.50cm}
\begin{center}
\includegraphics[width=9.50cm]{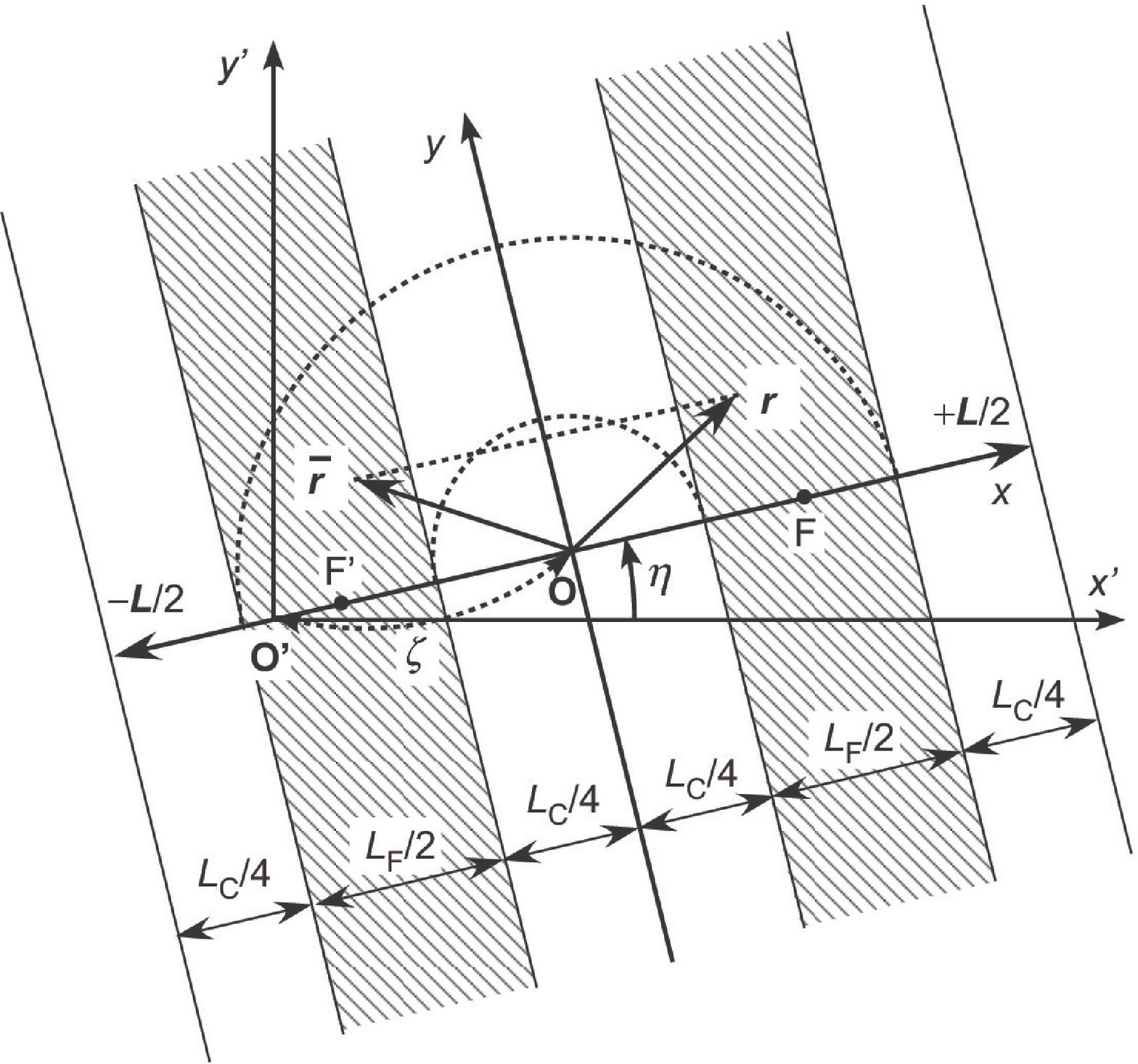}
(b)
\end{center}
\end{minipage}
\caption{%
\label{Fig:Flattened_CN}
(a) A schematic illustration of a collapsed carbon nanotubes and (b) its development map on graphene sheet.
In (b), the coordinates system $(x',y')$ and origin O' are fixed onto the graphene sheet and the coordinates system $(x,y)$ and origin O vary depending on the structure of a nanotube.
The flattened region is denoted by the shaded area.
}
\end{center}
\vspace{-0.250cm}
\end{figure*}
%
This paper is organized as follows: 
In Sect.\ \ref{Sec:Collapsed_Carbon_Nanotubes}, an effective potential of inter-wall interaction is derived in an effective-mass scheme.
In Sect.\ \ref{Sec:Weak_Inter-Wall_Coupling}, modification of band structure due to collapse is analyzed by perturbation of inter-wall interaction first for armchair and zigzag nanotube and its dependence on nanotube chirality and stacking in flattened region are discussed based on dominant terms.
Numerical results are shown in Sect.\ \ref{Sec:Numerical_Results} and a short summary is given in Sect.\ \ref{Sec:Summary_and_Conclusion}.
\par
%
\section{Collapsed Carbon Nanotubes} \label{Sec:Collapsed_Carbon_Nanotubes}
%
We consider a nanotube partially flattened as illustrated in Fig.\ \ref{Fig:Flattened_CN} (a).
The width of the flattened region is denoted by $L_F/2$ and that of the curved region by $L_C/2$.
We have
%
\begin{equation}
L_F + L_C = L ,
\end{equation}
%
where $L$ is the circumference.
Figure \ref{Fig:Flattened_CN} (b) shows the development map.
The tube is usually specified by chiral vector ${\bf L}$, corresponding to the circumference, i.e., $L=|{\bf L}|$.
The direction of ${\bf L}$ measured from the horizontal direction is called the chiral angle and denoted by $\eta$.
\par
%
In Fig.\ \ref{Fig:Flattened_CN} (b), the right hand side of the line passing through the point O at $(\zeta\cos\eta,\zeta\sin\eta)$ and perpendicular to ${\bf L}$ is folded down to form the lower half of the flattened nanotube.
The coordinate of the point in the lower side of the flattened region corresponding to point ${\bf r}$ in the upper side will be denoted by $\bar{\bf r}$.
Obviously, $\bar{\bf r}$ is given by the mirror reflection with respect to the line perpendicular to ${\bf L}$.
In the nanotube, we shall use the coordinates $(x,y)$ fixed onto the tube and therefore we have $\bar{\bf r}=(-x,y)$ for ${\bf r}=(x,y)$.
The coordinates of ${\bf r}$ and $\bar{\bf r}$ in the coordinates $(x',y')$ fixed onto the graphene sheet can be straightforwardly obtained.
\par
\begin{figure}
\begin{center}
\includegraphics[width=6.0cm]{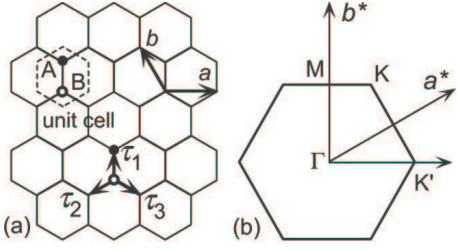}
\caption{%
\label{Fig:Lattice_Structure_and_Brillouin_Zone}
The lattice structure of graphene (a) and the first Brillouin zone (b).
The primitive translation vectors are denoted by ${\bf a}$ and ${\bf b}$ ($|{\bf a}|=|{\bf b}|=a$) and the vectors connecting nearest-neighbor atoms are denoted by $\vec\tau_1$, $\vec\tau_2$, and $\vec\tau_3$ in (a).
The reciprocal lattice vectors are denoted by ${\bf a}^*$ and ${\bf b}^*$, and ${\bf K}=(2\pi/a)(1/3,\,1/\sqrt3)$ and ${\bf K}'=(2\pi/a)(2/3,\,0)$ in (b).
}
\end{center}
\end{figure}

%
Figure \ref{Fig:Lattice_Structure_and_Brillouin_Zone} (a) shows the lattice structure of graphene, two primitive translation vectors ${\bf a}$ and ${\bf b}$, and three vectors $\vec\tau_l$ $(l=1,2,3)$ connecting nearest-neighbor atoms.
A unit cell contains two carbon atoms denoted by A and B.
In a tight-binding model, the wave function is written as
%
\begin{equation}
\psi({\bf r}) = \!\! \sum_{{\bf R}={\bf R}_A} \!\! \psi_A({\bf R}) \phi({\bf r}-{\bf R}) + \!\! \sum_{{\bf R}={\bf R}_B} \!\! \psi_B({\bf R}) \phi({\bf r}-{\bf R}) ,
\end{equation}
%
where $\phi({\bf r})$ denotes a $\pi$ orbital.
The amplitude $\psi$ at atomic sites ${\bf R}={\bf R}_A$ or ${\bf R}_B$ satisfies
%
\begin{eqnarray}
&& \!\!\!\!\!\! \varepsilon \psi_{A}({\bf R}_{A}) = - \gamma_0 \! \sum_{l=1}^3 \! \psi_{B}({\bf R}_{A} - \vec\tau_l) \nonumber \\
\noalign{\vspace{-0.20cm}}
&& \!\!\!\!\!\!\! + \! \sum_{{\bf R}_{A}'} V({\bf R}_{A},{\bf R}_{A}') \psi_{A}({\bf R}_{A}') + \! \sum_{{\bf R}_{B}'} V({\bf R}_{A},{\bf R}_{B}') \psi_{A}({\bf R}_{B}') , \quad \\
&& \!\!\!\!\!\! \varepsilon \psi_{B}({\bf R}_{B}) = - \gamma_0 \! \sum_{l=1}^3 \! \psi_{A}({\bf B}_{A} + \vec\tau_l) \nonumber \\
\noalign{\vspace{-0.20cm}}
&& \!\!\!\!\!\!\! + \! \sum_{{\bf R}_{A}'} V({\bf R}_{B},{\bf R}_{A}') \psi_{A}({\bf R}_{A}') + \sum_{{\bf R}_{B}'} V({\bf R}_{B},{\bf R}_{B}') \psi_{B}({\bf R}_{B}') , \quad
\label{Eq:TB_AB}
\end{eqnarray}
%
where $\gamma_0$ is the hopping integral between nearest-neighbor atoms within the wall and inter-wall hopping integral $V({\bf R},{\bf R}')$ is nonzero only when carbon atoms at sites ${\bf R}$ and ${\bf R}'$ are very closely located in the opposite side of the flattened region.
Since $\pi$ orbitals are symmetric within the wall, $V({\bf R},{\bf R}')$ is a function of $|{\bf R}-{\bf R}'|$ well inside the flattened region.
\par
%
%
In a monolayer graphene the conduction and valence bands consisting of $\pi$ orbitals cross at K and K' points of the Brillouin zone shown in Fig.\ \ref{Fig:Lattice_Structure_and_Brillouin_Zone} (b), where the Fermi level is located.\cite{Wallace_1947a,Painter_and_Ellis_1970a}
For states in the vicinity of the Fermi level $\varepsilon=0$, the amplitudes are written as
%
\begin{eqnarray}
&& \!\!\!\! \psi_{A}({\bf R}_{A}) = e^{i {\bf K}\cdot{\bf R}_{A}} F_{A}^K({\bf R}_{A}) + e^{+i \eta} e^{i {\bf K}' \cdot{\bf R}_{A}} F_{A}^{K'}({\bf R}_{A}) , \\
&& \!\!\!\! \psi_{B}({\bf R}_{B}) = - \omega e^{+ i \eta} e^{i {\bf K}\cdot{\bf R}_{B}} F_{B}^K({\bf R}_{B}) + e^{i {\bf K}'\cdot{\bf R}_{B}} F_{B}^{K'}({\bf R}_{B}) , \nonumber \\
\noalign{\vspace{-0.10cm}}
\end{eqnarray}
%
with $\omega=\exp(2\pi i/3)$.\cite{Ando_2005a_and_References}
Envelope functions $F_{A}^K$, $F_{B}^K$, $F_{A}^{K'}$, and $F_{B}^{K'}$ are assumed to be slowly varying in the scale of the lattice constant.
The effective-mass approximation is valid and well reproduces electronic properties as well as the band structure for energy range given by $|\varepsilon|\ll 3\gamma_0$.\cite{Ajiki_and_Ando_1993a,Ajiki_and_Ando_1996a,Ando_2005a_and_References}
\par
%
In the absence of inter-wall interactions, the envelope functions for the K point satisfy
%
\begin{eqnarray}
\hat{\cal H}^K(\hat{\bf k}) {\bf F}^K({\bf r}) \!\! & = & \!\! \varepsilon {\bf F}^K({\bf r}) , \\
\hat{\cal H}^K(\hat{\bf k}) \!\! & = & \!\! \gamma \pmatrix{ 0 & \hat k_- \cr \hat k_+ & 0 \cr } , \\
{\bf F}^K({\bf r}) \!\! & = & \!\! \pmatrix{ F_A^K({\bf r}) \cr \noalign{\vspace{0.10cm}} F_B^K({\bf r}) \cr } .
\end{eqnarray}
%
where $\hat k_\pm=\hat k_x\pm i \hat k_y$, $\hat{\bf k}=-i\vec\nabla$, and $\gamma=\sqrt{3}a\gamma_0/2$ is the band parameter with lattice constant $a=0.246$ nm.
Here, wave vector ${\bf k}$ is measured from the K point denoted by ${\bf K}$.
For the K' point, we should exchange $\hat k_+$ and $\hat k_-$ in the Hamiltonian, i.e.,
%
\begin{equation}
\hat{\cal H}^{K'}(\hat{\bf k}) = \gamma \pmatrix{ 0 & \hat k_+ \cr \hat k_- & 0 \cr } ,
\end{equation}
%
where ${\bf k}$ is measured from the K' point, i.e., ${\bf K}'$.
\par
%
We shall construct a nanotube in such a way that the hexagon at ${\bf L}=n_a {\bf a} + n_b {\bf b}$ with integers $n_a$ and $n_b$ is rolled onto the origin.
For translation ${\bf r}\!\rightarrow\!{\bf r}+{\bf L}$, the Bloch function at the K and K' points acquires the phase
%
\begin{eqnarray}
\exp( i {\bf K} \cdot {\bf L}) \!\! & = & \!\! \exp\Big( + {2\pi i \nu \over 3} \Big) , \\
\exp( i {\bf K}' \cdot {\bf L}) \!\! & = & \!\! \exp\Big( - {2\pi i \nu \over 3} \Big) ,
\end{eqnarray}
%
where $\nu=0$ or $\pm1$, determined by
%
\begin{equation}
n_a + n_b = 3 N + \nu ,
\label{Eq:nu}
\end{equation}
%
with integer $N$.
Correspondingly, the boundary conditions for ${\bf F({\bf r})}$ are given by
%
\begin{eqnarray}
{\bf F}^K({\bf r}+{\bf L}) = \exp\Big( - {2\pi i \nu \over 3} \Big) {\bf F}^K({\bf r}) , \\
{\bf F}^{K'}({\bf r}+{\bf L}) = \exp\Big( + {2\pi i \nu \over 3} \Big) {\bf F}^{K'}({\bf r}) ,
\end{eqnarray}
%
\par
%
Let ${\bf T}$ be the primitive lattice translation vector in the axis direction,
%
\begin{equation}
{\bf T} = m_a {\bf a} + m_b {\bf b} ,
\label{Eq:TranslationVector}
\end{equation}
%
with integers $m_a$ and $m_b$.
We have
%
\begin{equation}
p m_a = n_a - 2 n_b , \qquad p m_b = 2 n_a - n_b ,
\end{equation}
%
where $p$ is the greatest common divisor of $n_a-2n_b$ and $2n_a-n_b$.
For translation ${\bf r}\!\rightarrow\!{\bf r}+{\bf T}$, the Bloch function at the K and K' points acquires the phase
%
\begin{eqnarray}
\exp( i {\bf K} \cdot {\bf T}) & \!\! = \!\! & \exp\Big( + {2\pi i \mu \over 3} \Big) , \\
\exp( i {\bf K}' \cdot {\bf T}) & \!\! = \!\! & \exp\Big( - {2\pi i \mu \over 3} \Big) ,
\end{eqnarray}
%
where $\mu=0$ or $\pm1$, determined by
%
\begin{equation}
m_a + m_b = 3 M + \mu ,
\label{Eq:mu}
\end{equation}
%
with integer $M$.
This shows that the K and K' points are mapped onto $k_\mu^K$ and $k_\mu^{K'}$, respectively, with
%
\begin{equation}
k_\mu^K \equiv + {2\pi \mu \over 3 T} , \quad k_\mu^{K'} \equiv - {2\pi \mu \over 3 T} .
\end{equation}
%
within the one-dimensional first Brillouin zone $[-\pi/T,\allowbreak +\pi/T]$, with $T=|{\bf T}|$.
\par
\begin{figure*}
\begin{center}
\includegraphics[height=6.0cm]{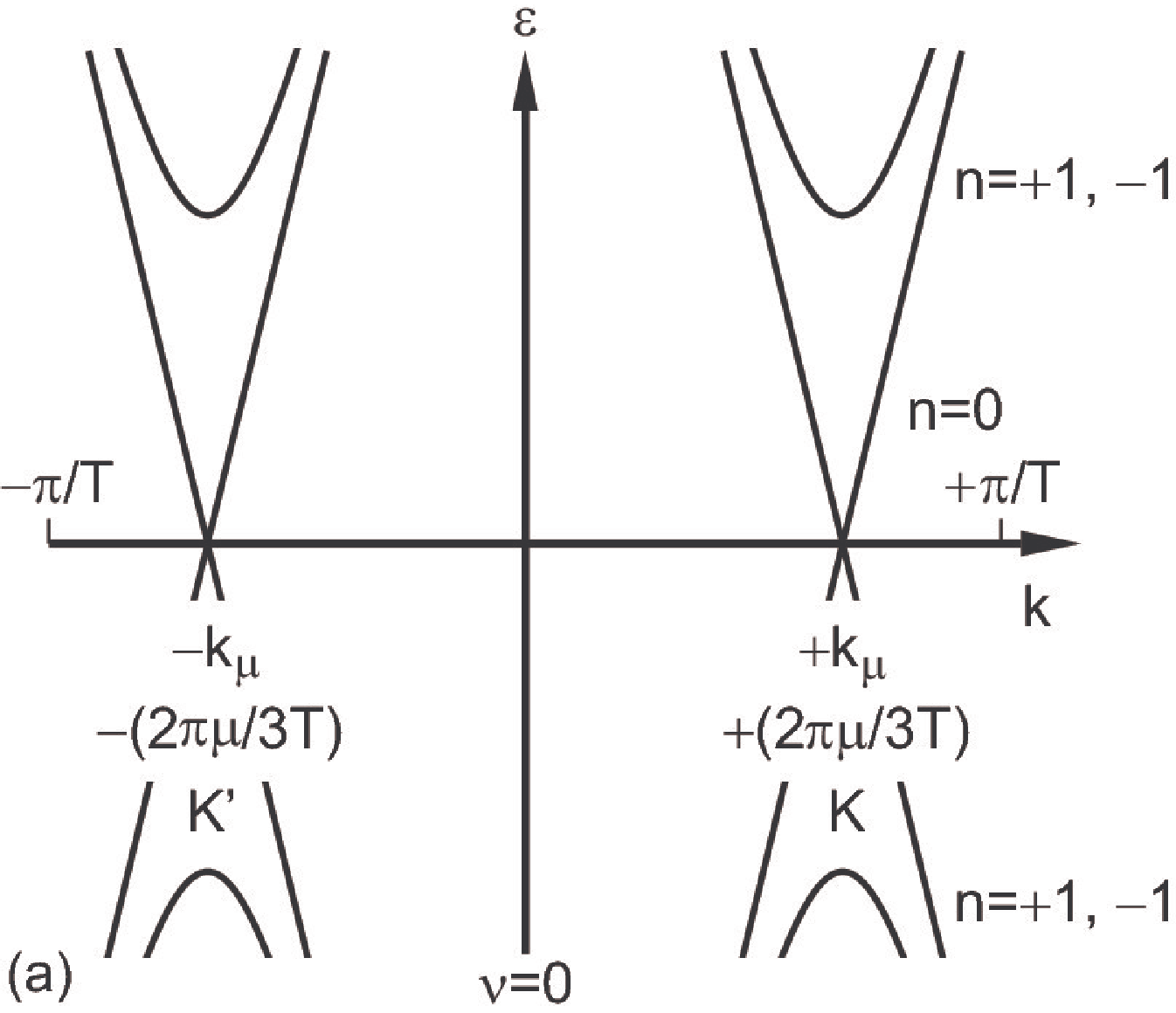}
\qquad\quad
\includegraphics[height=6.0cm]{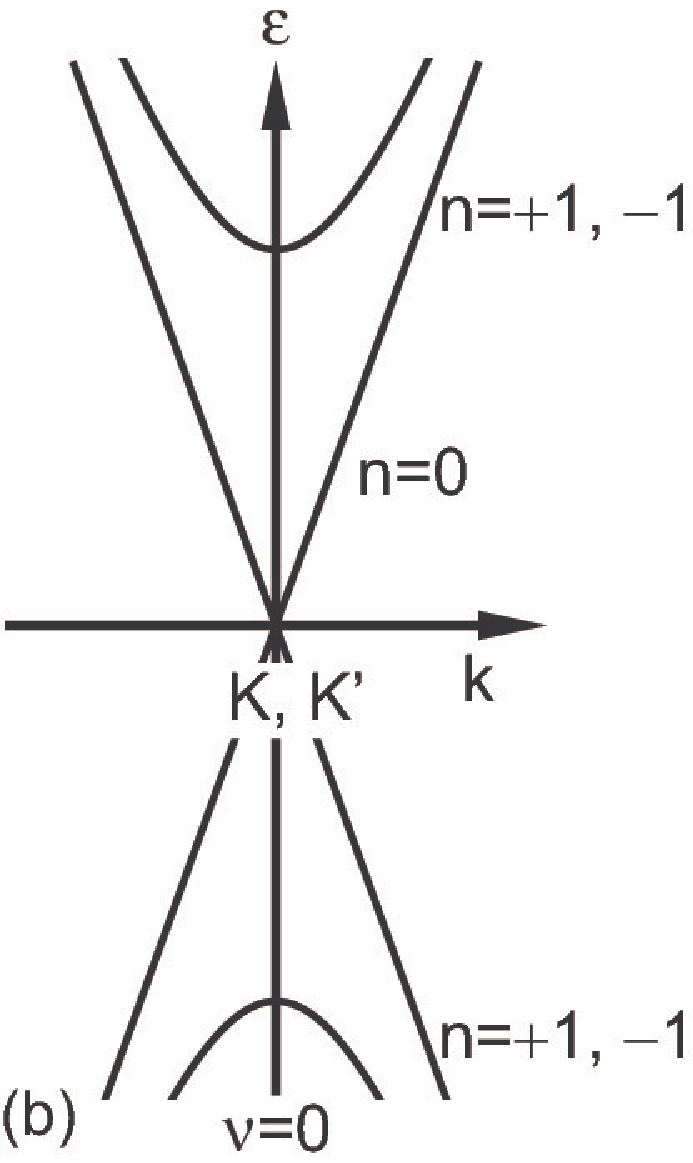}
\qquad\quad
\includegraphics[height=6.0cm]{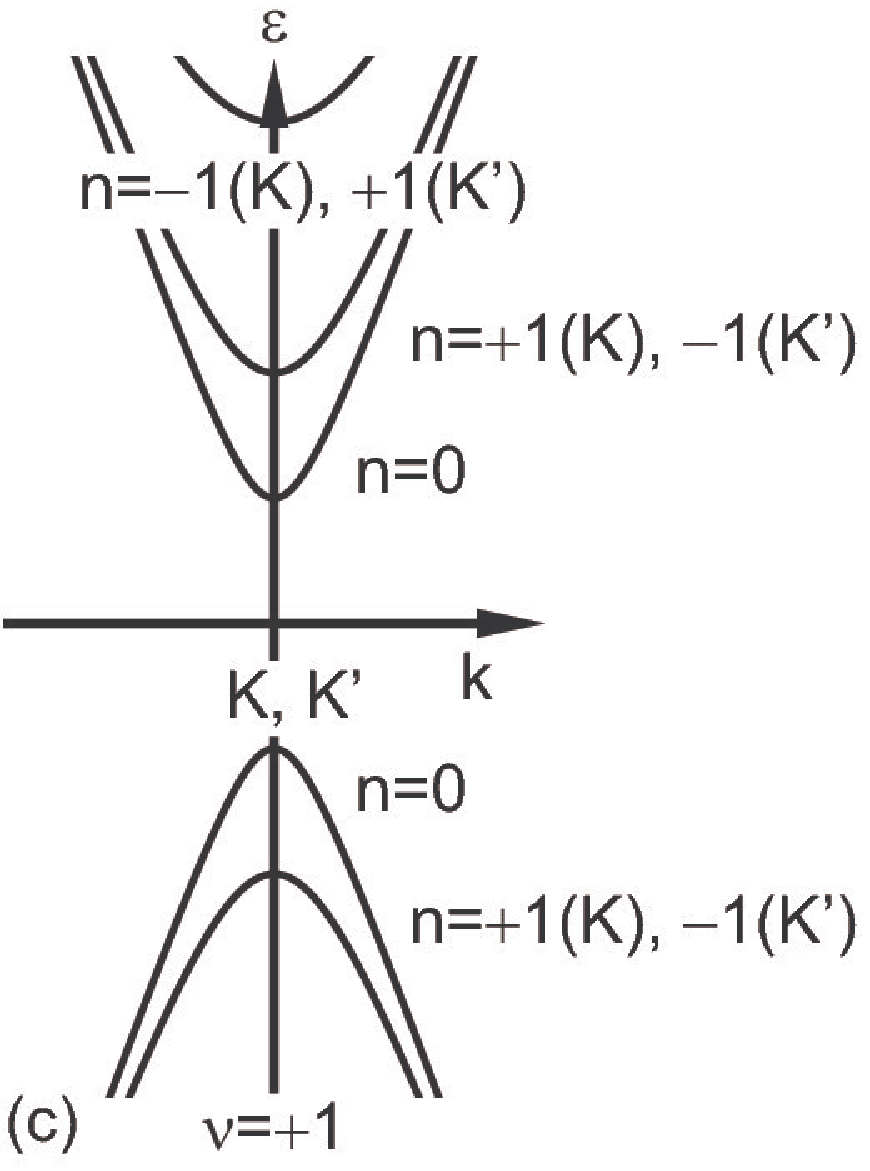}
\caption{%
\label{Fig:Nanotube_Bands}
Schematic illustration of energy bands for (a) an armchair tube $k_\mu\!\neq\!0$, and (b) metallic and (c) semiconducting zigzag nanotubes with $k_\mu=0$.
}
\end{center}
\end{figure*}

%
For the K point, the energies and corresponding wave functions are given by\cite{Ajiki_and_Ando_1993a,Ajiki_and_Ando_1996a,Ando_2005a_and_References}
%
\begin{eqnarray}
\varepsilon_{ns}^K(k) \!\! & = & \!\! s \gamma \sqrt{\kappa_\nu^K(n)^2 + (k - k_\mu^K)^2 } , \\
{\bf F}_{nks}^K({\bf r}) \!\! & = & \!\! {1\over \sqrt{AL}} {\bf F}_{nks}^K \exp\big[ i \kappa_\nu^K(n) x + i (k - k_\mu^K) y \big] , \quad
\end{eqnarray}
%
with
%
\begin{eqnarray}
\kappa_\nu^K(n) \!\! & = & \!\! {2\pi \over L} \Big( n - {\nu\over 3} \Big) , \\
{\bf F}_{nks}^K \!\! & = & \!\! {1\over \sqrt2} \pmatrix{ b_{\mu,\nu}^K(n,k) \cr s \cr } , \\
b_{\mu\nu}^K(n,k) \!\! & = & \!\! { \kappa_\nu^K(n) - i (k - k_\mu^K) \over \sqrt{\kappa_\nu(n)^2 + (k - k_\mu^K)^2} } ,
\end{eqnarray}
%
where $k$ is the wave vector in the axis direction, measured from the center of the one-dimensional Brillouin zone, $n$ is an integer, $s=+1$ and $-1$ for the conduction and valence band, respectively, and $A$ is the tube length.
For the K' point, we should replace $\nu$ with $-\nu$, $\mu$ with $-\mu$, and $b_{\mu,\nu}^K(n,k)$ with $b_{\mu,\nu}^{K'}(n,k)=b_{-\mu,-\nu}^{K}(n,k)^*$.
The band structure is illustrated in Fig.\ \ref{Fig:Nanotube_Bands} in the vicinity of $\varepsilon=0$.
\par
%
%
In the presence of inter-wall coupling, the envelope functions satisfy the Schr\"odinger equation,
%
\begin{equation}
\hat{\cal H}(\hat{\bf k}) {\bf F}({\bf r}) + \int d {\bf r}' \hat{\cal V}({\bf r},{\bf r}') {\bf F}({\bf r}') = \varepsilon {\bf F}({\bf r}) ,
\label{Eq:Schrodinger_equation}
\end{equation}
%
with
%
\begin{eqnarray}
\hat {\cal H}(\hat{\bf k}) \!\! & = & \!\! \pmatrix{ \hat{\cal H}^K(\hat{\bf k}) & 0 \cr 0 & \hat{\cal H}^{K'}(\hat{\bf k}) \cr } , \\
{\bf F}({\bf r}) \!\! & = & \!\! \pmatrix{ {\bf F}^K({\bf r}) \cr {\bf F}^{K'}({\bf r}) \cr } . 
\end{eqnarray}
%
The (4,4) matrix effective-potential of inter-wall interaction is given by
%
\begin{equation}
\hat{\cal V}({\bf r},{\bf r}') = \hat V({\bf r}) \delta({\bf r}'-\bar{\bf r}) ,
\end{equation}
%
with
%
\begin{equation}
\hat V({\bf r}) = \pmatrix{\hat V^{KK}({\bf r}) & \hat V^{KK'}({\bf r}) \cr \hat V^{K'K}({\bf r}) & \hat V^{K'K'}({\bf r}) \cr } ,
\end{equation}
%
where $\hat V^{KK}({\bf r})$, etc.\ are (2,2) matrices given by
%
\begin{equation}
\hat V^{KK}({\bf r}) = \pmatrix{V_{AA}^{KK}({\bf r}) & V_{AB}^{KK}({\bf r}) \cr \noalign{\vspace{0.10cm}} V_{BA}^{KK}({\bf r}) & V_{BB}^{KK}({\bf r}) \cr } ,
\end{equation}
%
etc.
Explicit expressions for the effective potential are more easily written down in terms of
%
\begin{equation}
\tilde V({\bf r}) = \pmatrix{\tilde V_{AA}({\bf r}) & \tilde V_{AB}({\bf r}) \cr \tilde V_{BA}({\bf r}) & \tilde V_{BB}({\bf r}) \cr } ,
\end{equation}
%
where $\tilde V_{AA}({\bf r})$, etc.\ are (2,2) matrices given by
%
\begin{equation}
\tilde V_{AA}({\bf r}) = \pmatrix{ \tilde V_{AA}^{KK}({\bf r}) & \tilde V_{AA}^{KK'}({\bf r}) \cr \noalign{\vspace{0.10cm}} \tilde V_{AA}^{K'K}({\bf r}) & \tilde V_{AA}^{K'K'}({\bf r}) \cr } ,
\end{equation}
%
etc.
We can obtain $\tilde V$ from $\hat V({\bf r})$ by a simple unitary transformation.
\par
%
Then, the effective potential of inter-wall interaction is explicitly given by
%
\begin{eqnarray}
&& \!\!\!\! \tilde V_{AA}({\bf r}) = \!\!\! \sum_{{\bf R}_{A},{\bf R}_{A}'} \!\! {1\over 2} \big[ g({\bf r}-{\bf R}_{A}) + g(\bar{\bf r}-{\bf R}_{A}') \big] V({\bf R}_{A},{\bf R}_{A}') \quad \nonumber \\
&& \!\!\!\! \times \! \pmatrix{
e^{i {\bf K}\cdot ({\bf R}'_{A}-{\bf R}_{A})} &
 e^{i \eta}e^{i ({\bf K}'\cdot{\bf R}'_{A}-{\bf K}\cdot{\bf R}_{A})} \cr 
 e^{-i \eta}e^{i ({\bf K}\cdot{\bf R}'_{A}-{\bf K}'\cdot{\bf R}_{A})} & 
e^{i {\bf K}'\cdot ({\bf R}'_{A}-{\bf R}_{A})} \cr } \! , \label{Eq:VAA} \\
&& \!\!\!\! \tilde V_{AB}({\bf r}) = \!\!\! \sum_{{\bf R}_{A},{\bf R}_{B}'} \!\! {1\over 2} \big[ g({\bf r}-{\bf R}_{A}) + g(\bar{\bf r}-{\bf R}_{B}') \big] V({\bf R}_{A},{\bf R}_{B}') \nonumber \\
&& \!\!\!\! \times \! \pmatrix{
 -\omega e^{i \eta}e^{i {\bf K}\cdot ({\bf R}'_{B}-{\bf R}_{A})} &
e^{i ({\bf K}'\cdot{\bf R}'_{B}-{\bf K}\cdot{\bf R}_{A})}\cr
-\omega e^{i ({\bf K}\cdot{\bf R}'_{B}-{\bf K}'\cdot{\bf R}_{A})} & 
e^{-i \eta}e^{i {\bf K}'\cdot ({\bf R}'_{B}-{\bf R}_{A})} \cr } \! , \label{Eq:VAB} \\
&& \!\!\!\! \tilde V_{BA}({\bf r}) = \!\!\! \sum_{{\bf R}_{A},{\bf R}_{A}'} \!\! {1\over 2} \big[ g({\bf r}-{\bf R}_{B}) + g(\bar{\bf r}-{\bf R}_{A}') \big] V({\bf R}_{B},{\bf R}_{A}') \nonumber \\
&& \!\!\!\! \times \! \pmatrix{
\! -\omega^{-1} e^{-i \eta}e^{i {\bf K}\cdot ({\bf R}'_{A}-{\bf R}_{B})} &
 \!\! -\omega^{-1} e^{i ({\bf K}'\cdot{\bf R}'_{A}-{\bf K}\cdot{\bf R}_{B})} \! \cr
e^{i ({\bf K}\cdot{\bf R}'_{A}-{\bf K}'\cdot{\bf R}_{B})} & 
e^{i \eta} e^{i {\bf K}'\cdot ({\bf R}'_{A}-{\bf R}_{B})} \cr } \! , \label{Eq:VBA} \\
&& \!\!\!\! \tilde V_{BB}({\bf r}) = \!\!\! \sum_{{\bf R}_{B},{\bf R}_{B}'} \!\! {1\over 2} \big[ g({\bf r}-{\bf R}_{B}) + g(\bar{\bf r}-{\bf R}_{B}') \big] V({\bf R}_{B},{\bf R}_{B}') \nonumber \\
&& \!\!\!\! \times \! \pmatrix{
\!\!\! e^{i {\bf K}\cdot ({\bf R}'_{B}-{\bf R}_{B})} & \!\!\! \!\!\! -\omega^{-1} e^{-i \eta} e^{i ({\bf K}'\cdot{\bf R}'_{B}-{\bf K}\cdot{\bf R}_{B})} \cr
- \omega e^{i \eta} e^{i ({\bf K}\cdot{\bf R}'_{B}-{\bf K}'\cdot{\bf R}_{B})} \!\!\!\!\!\!  & e^{i {\bf K}'\cdot ({\bf R}'_{B}-{\bf R}_{B})} \!\!\! \cr } \! . \label{Eq:VBB} 
\end{eqnarray}
%
We should note that ${\bf R}_A$ and ${\bf R}_B$ and also ${\bf K}$ and ${\bf K}'$ are in the coordinate system $x'y'$ fixed onto the development map and ${\bf r}$ and $\bar{\bf r}$ are in the coordinate system $xy$ fixed onto carbon nanotubes.
Thus, ${\bf r}$ and $\bar{\bf r}$ should be converted into the $x'y'$ system in the above equations.
\par
%
We have introduced a smoothing function $g({\bf r})$ which varies smoothly in the range $|{\bf r}| \alt a$ and decays rapidly and vanishes for $|{\bf r}|\gg a$.\cite{Ando_2005a_and_References}
It should satisfy the conditions:
%
\begin{eqnarray}
&& \!\!\!\! \sum_{{\bf R}_A} g({\bf r} - {\bf R}_A) = \sum_{{\bf R}_B} g({\bf r} - {\bf R}_B) = 1 , \\
&& \!\!\!\! \int d {\bf r} \, g({\bf r} - {\bf R}_A) = \int d {\bf r} \, g({\bf r} - {\bf R}_B) = \Omega_0 ,
\end{eqnarray}
%
where $\Omega_0$ is the area of a unit cell given by $\Omega_0=\sqrt3 a^2/2$.
The function $g({\bf r} - {\bf R})$ can be replaced by a delta function when it is multiplied by a function such as ${\bf F}({\bf r})$ varying smoothly in the scale of the lattice constant, i.e., $g({\bf r}-{\bf R}) \approx \Omega_0 \delta({\bf r}-{\bf R})$.
\par
%
The effective potential satisfies
%
\begin{equation}
\hat V(\bar{\bf r}) = \hat V({\bf r})^\dagger , 
\label{Eq:Symmetry_of_Interlayer_Interaction}
\end{equation}
%
and therefore
%
\begin{equation}
\hat{\cal V}({\bf r},{\bf r}')^\dagger = \hat V(\bar{\bf r}) \delta({\bf r}'-\bar{\bf r}) = \hat V({\bf r}') \delta({\bf r}-\bar{\bf r}') = \hat{\cal V}({\bf r}',{\bf r}) ,
\end{equation}
%
which insures that $\hat{\cal V}({\bf r},{\bf r}')$ is a Hermitian operator.
The Hamiltonian should satisfy the time reversal invariance under operation given by\cite{Suzuura_and_Ando_2002b,Ando_2006c}
%
\begin{equation}
{\bf F}^T = e^{- i \psi} \pmatrix{0 & \sigma_z \cr \sigma_z & 0 \cr } {\bf F}^* .
\end{equation}
%
where $\sigma_x$ is the Pauli spin matrix and $\psi$ is an arbitrary phase factor.
Thus, the inter-wall potential $\hat V({\bf r})$ should satisfy 
%
\begin{equation}
\pmatrix{ 0 & \sigma_z \cr \sigma_z & 0 \cr } \hat V({\bf r})^* \pmatrix{ 0 & \sigma_z \cr \sigma_z & 0 \cr } = \hat V({\bf r}) ,
\label{Eq:Time_Reversal_Symmetry_of_Interlayer_Potential}
\end{equation}
%
or
%
\begin{equation}
\pmatrix{ \sigma_x & 0 \cr 0 & - \sigma_x \cr } \tilde V({\bf r})^* \pmatrix{ \sigma_x & 0 \cr 0 & - \sigma_x \cr  } = \tilde V({\bf r}) .
\label{Eq:Time_Reversal_Symmetry_of_Interlayer_Potential:_2}
\end{equation}
%
\par
%
Further, the effective potential has the following translational properties:
%
\begin{eqnarray}
\hat V^{KK}({\bf r}+{\bf L}) \!\! & = & \!\! \hat V^{KK}({\bf r}) \, e^{-4\pi i \nu/3} , \nonumber \\
\hat V^{KK'}({\bf r}+{\bf L}) \!\! & = & \!\! \hat V^{KK'}({\bf r}) , \nonumber \\
\noalign{\vspace{-0.250cm}}
\\
\noalign{\vspace{-0.250cm}}
\hat V^{K'K}({\bf r}+{\bf L}) \!\! & = & \!\! \hat V^{K'K}({\bf r}) , \nonumber \\
\hat V^{K'K'}({\bf r}+{\bf L}) \!\! & = & \!\! \hat V^{K'K'}({\bf r}) \, e^{+4\pi i \nu/3} , \nonumber
\end{eqnarray}
%
and
%
\begin{eqnarray}
\hat V^{KK}({\bf r}+{\bf T}) \!\! & = & \!\! \hat V^{KK}({\bf r}) , \nonumber \\
\hat V^{KK'}({\bf r}+{\bf T}) \!\! & = & \!\! \hat V^{KK'}({\bf r}) \, e^{-4\pi i \mu/3} , \nonumber \\
\noalign{\vspace{-0.250cm}}
\\
\noalign{\vspace{-0.250cm}}
\hat V^{K'K}({\bf r}+{\bf T}) \!\! & = & \!\! \hat V^{K'K}({\bf r}) \, e^{+4\pi i \mu/3} , \nonumber \\
\hat V^{K'K'}({\bf r}+{\bf T}) \!\! & = & \!\! \hat V^{K'K'}({\bf r}) . \nonumber
\end{eqnarray}
%
Therefore, it can be expanded into a Fourier series, such that
%
\begin{eqnarray}
\hat V^{KK}({\bf r}) \!\! & = & \!\! \sum_{n,m} \hat V^{KK}_{n,m} \! \exp \! \Big[{2\pi i \over L}\Big(n - {2\nu \over 3}\Big) x + {2\pi i m\over T} y \Big] , \nonumber \\
\hat V^{KK'}({\bf r}) \!\! & = & \!\! \sum_{n,m} \hat V^{KK'}_{n,m} \! \exp \! \Big[{2\pi i n\over L} x + {2\pi i \over T} \Big( m - {2\mu\over 3} \Big) y \Big] , \nonumber \\
\noalign{\vspace{-0.3750cm}}
\\
\noalign{\vspace{-0.150cm}}
\hat V^{K'K}({\bf r}) \!\! & = & \!\! \sum_{n,m} \hat V^{K'K}_{n,m} \! \exp \! \Big[{2\pi i n\over L} x + {2\pi i \over T} \Big( m + {2\mu\over 3} \Big) y \Big] , \quad \nonumber \\
\hat V^{K'K'}({\bf r}) \!\! & = & \!\! \sum_{n,m} \hat V^{K'K'}_{n,m} \! \exp \! \Big[{2\pi i \over L}\Big(n + {2\nu \over 3}\Big) x + {2\pi i m\over T} y \Big] , \nonumber
\end{eqnarray}
%
with integers $n$ and $m$.
This shows that the inter-wall coupling gives rise to interactions among bands with same $k$ value in the one-dimensional Brillouin zone.
\par
%
We shall expand the wave functions in terms of those of the corresponding cylindrical nanotube:
%
\begin{eqnarray}
&& \!\!\!\! {\bf F}({\bf r}) = {1\over \sqrt{AL}} \sum_{m,n} \exp[i (k+G_m) y] \nonumber \\
&& \!\!\!\! \times \pmatrix{ \exp[i \kappa_\nu^{K}(n)x-i k_\mu^K y] \!\!\!\!\!\!\!\!\!\! & 0 \!\!\!\!\!\!\!\! \cr \noalign{\smallskip} \!\!\!\!\!\!\!\! 0 & \!\!\!\!\!\!\!\!\!\! \exp[i \kappa_\nu^{K'}(n)x-i k_\mu^{K'} y] \cr } {\bf F}_{nm} , \qquad
\end{eqnarray}
%
with $G_m=2\pi m/T$.
Then, we have
%
\begin{equation}
\hat {\cal H}[n,k+G_m] {\bf F}_{nm} + \sum_{n'm'} \hat V_{n+n',m-m'} {\bf F}_{n'm'} = \varepsilon {\bf F}_{nm} ,
\end{equation}
%
with
%
\begin{equation}
\hat {\cal H}[n,k] = \pmatrix {\hat{\cal H}^{K}[\kappa_\nu^{K}(n),k-k_\mu^K] \!\!\!\!\!\! & 0 \!\!\!\!\!\! \cr \noalign{\smallskip} \!\!\!\!\!\! 0 & \!\!\!\!\!\! \hat{\cal H}^{K'}[\kappa_\nu^{K'}(n),k-k_\mu^{K'}] \cr} \! ,
\end{equation}
%
and
%
\begin{equation}
\hat V_{n+n',m-m'} = \pmatrix{ \hat V_{n+n',m-m'}^{KK} & \hat V_{n+n',m-m'}^{KK'} \cr \noalign{\smallskip} \hat V_{n+n',m-m'}^{K'K} & \hat V_{n+n',m-m'}^{K'K'} \cr } .
\end{equation}
%
\par
%
\begin{figure*}
\begin{center}
\includegraphics[width=15.0cm]{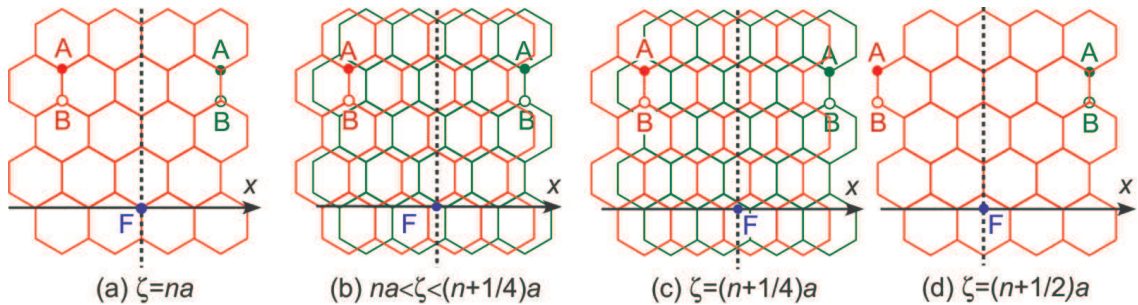}
\caption{%
\label{Fig:Structure_of_Flattened_Zigzag_Nanotube}
Some examples of the structure of the double-wall region of a flattened zigzag nanotube.
The A and B sites remain the same, but the K and K' points are exchanged between the upper (red) region and the lower (green) region.
(a) $\zeta=na$ corresponding to an AA stacked bilayer, (b) $na<\zeta(n+{1 \over 4})a$, (c) $\zeta=(n+{1\over4})a$, (d) $\zeta=(n+{1\over 2})a$ corresponding to another AA stacked bilayer.
}
\end{center}
\vspace{-0.250cm}
\end{figure*}
%
For actual numerical calculations, the inter-wall hopping integral $V({\bf R}_1,{\bf R}_2)$ is chosen as\cite{Slater_and_Koster_1954a,Spain_1973a,Dresselhaus_et_al_1988a,Mintmire_and_White_1995a,Nakanishi_and_Ando_2001a,Uryu_2004a,Moon_and_Koshino_2012a,Moon_and_Koshino_2013a}
%
\begin{eqnarray}
\!\!\!\! && V({\bf R}_1,{\bf R}_2) = - \Big[ \alpha {\gamma_1 \over |{\bf t}|^2 } \exp \! \Big( \! - {|{\bf t}| - c/2 \over \delta } \Big) ({\bf p}_1 \! \cdot \! {\bf t}) ({\bf p}_2 \! \cdot \! {\bf t}) \nonumber \\
\!\!\!\! &&- \gamma_0 \exp \! \Big( \! - {|{\bf t}| - b \over \delta } \Big) [ ({\bf p}_1 \! \cdot \! {\bf u}) ({\bf p}_2 \! \cdot \! {\bf u}) + ({\bf p}_1 \! \cdot \! {\bf v}) ({\bf p}_2 \! \cdot \! {\bf v}) ] \Big] , \qquad 
\end{eqnarray}
%
where $b$ is the distance between neighboring carbons in graphene, i.e., $b=a/\sqrt3$, $c$ the lattice constant along the $c$ axis in graphite given by $c/a=2.72$, and $\delta$ the decay rate of the $\pi$ orbital.
Further, $\gamma_1$ is the hopping integral between nearest-neighbor sites of neighboring layers in graphite.
Vectors, ${\bf p}_1$ and ${\bf p}_2$ are unit vectors directed along $\pi$ orbitals at ${\bf R}_1$ and at ${\bf R}_2$, respectively, ${\bf t}$ is a vector connecting the two sites, and ${\bf u}$ and ${\bf v}$ are unit vectors perpendicular to ${\bf t}$ and to each other.
In the following numerical calculations, we use parameters $\gamma_0=2.7$ eV, $\gamma_1=0.4$ eV, $\delta/a=0.185$, and $\alpha=1.4$.
\par
%
The negative sign appearing in $V({\bf R}_1,{\bf R}_2)$ is due to the fact that the $\pi$ orbitals in the top and bottom sides of the nanotube have signs opposite to each other because of the tube geometry.
Further, we choose the following smoothing function:
%
\begin{equation}
g({\bf r}) = {\Omega_0 \over \pi d^2 } \exp\Big( - {{\bf r}^2 \over d^2 } \Big) ,
\end{equation}
%
with smoothing length $d$.
This $d$ is of the order of lattice constant $a$, but can be regarded as zero in the scale of the effective-mass approximation.
In the following, we choose $2\,\lsim\,d/a\,\lsim\,5$, for which the results are independent of this parameter.
\par
%
As shown by molecular dynamics simulations, coupling in the flattened region is slowly turned on in the vicinity of its edge.\cite{Zhang_et_al_2006b}
When the coupling suddenly appears at a boundary, extra coupling terms may appear, being strongly localized at edges.
In order to avoid such unphysical effects, we multiply the inter-wall hopping by the following function:
%
\begin{equation}
\theta(x-x_{\rm edge},\Delta_{\rm edge}) = {1\over 2} \Big[ 1 - {\rm erf}\Big({x - x_{\rm edge} \over \Delta_{\rm edge}}\Big) \Big] ,
\label{Eq:Delta_edge}
\end{equation}
%
with ${\rm erf}(t)$ being the error function defined by
%
\begin{equation}
{\rm erf}(t) = {2\over \sqrt\pi} \int_0^t e^{-s^2} d s .
\end{equation}
%
The parameter $\Delta_{\rm edge}$ describes the width of the region where coupling increases from zero to the value well in the flattened region.
Actual calculations show that an extra effective potential localized at edges appears  for $\Delta_{\rm edge}/a\ll 1$, but turns out to be negligibly small except in very narrow wires and therefore can safely be neglected for thick wires with collapsed structures.
\par
%
\section{Weak Inter-Wall Coupling} \label{Sec:Weak_Inter-Wall_Coupling}
\subsection{Inter-Wall Potential in Zigzag and Armchair} \label{Ssc:Inter-Wall_Potential}
%
\begin{figure*}
\begin{center}
\includegraphics[width=15.0cm]{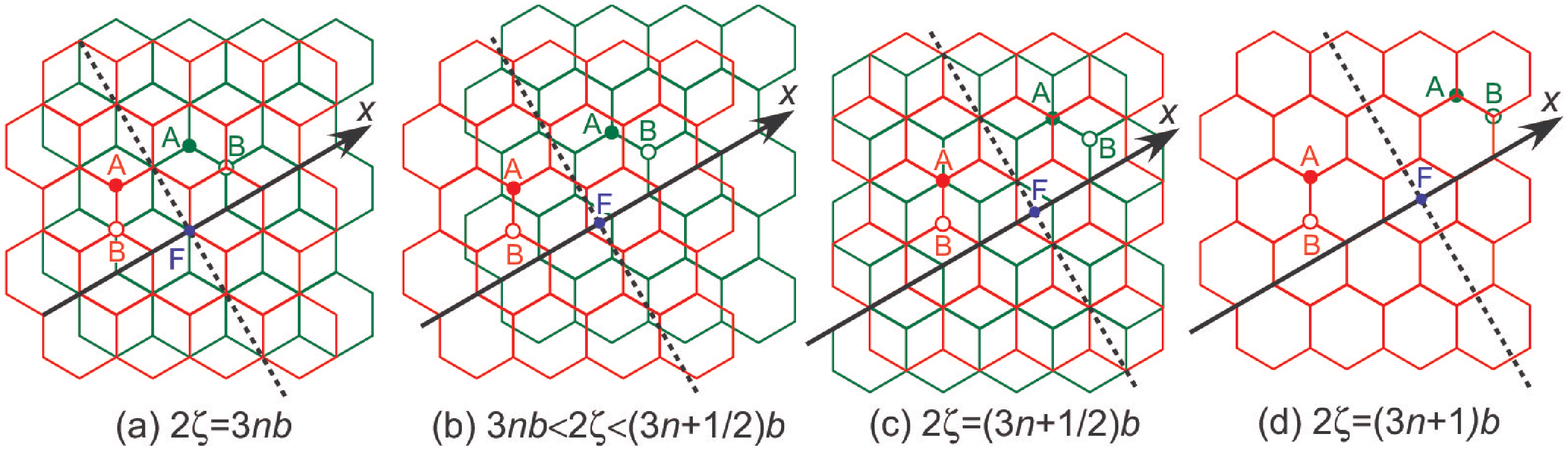}
\caption{%
\label{Fig:Structure_of_Flattened_Armchair_Nanotube}
Some examples of the structure of the double-wall region of a flattened armchair nanotube.
The K and K' points remain the same, but A and B sites are exchanged between the upper (red) region and the lower (green) region.
(a) $2\zeta=3nb$ corresponding to AB stacked bilayer, (b) $3nb<2\zeta<(3n+{1\over2})b$, (c) $2\zeta=(3n+{1\over2})b$ corresponding to AB stacking, (d) $2\zeta=(3n+1)b$ corresponding to AA stacking, with distance $b=a/\sqrt3$ between neighboring carbon atoms within the plane.
}
\end{center}
\vspace{-0.50cm}
\end{figure*}
%
Figure \ref{Fig:Structure_of_Flattened_Zigzag_Nanotube} shows some examples of the structure of the flattened region in a collapsed zigzag tube ($\eta=0$).
In zigzag tubes, A and B sublattices remain the same, but intra-valley components of inter-wall potential identically vanish, because the K point is mapped onto the K' point and the K' point onto the K point by the mirror reflection with respect to a line parallel to the axis.
The effective potential in the flattened bilayer region becomes independent of position and is periodic as a function of $\zeta$ with period $a/2$.
\par
%
We have in general
%
\begin{eqnarray}
\tilde V_{AA} \!\! & = & \!\! - \tilde V_{BB} = - v_1 \pmatrix{ 0 & e^{ i \varphi_1} \cr e^{- i \varphi_1} & 0 \cr } , \nonumber \\
\noalign{\vskip-0.250cm}
\\
\noalign{\vskip-0.250cm}
\tilde V_{AB} \!\! & = & \!\! - \tilde V_{BA} = - v_2 \pmatrix{ 0 & - e^{ i \varphi_2} \cr e^{- i \varphi_2} & 0 \cr } , \nonumber
\end{eqnarray}
%
with real coefficients $v_1$ and $v_2$ and phases $\varphi_1$ and $\varphi_2$ varying with $\zeta$.
Actually, difference $\varphi_2-\varphi_1$ is a relevant parameter changing the band structure, because a relative phase difference between the wave functions associated with the K and K' point can be chosen arbitrarily and is  not important.
\par
%
For the above potential, the band structure in the bilayer region generally consists of two cone-like bands with crossing points displaced in the wave vector space and have different energies.
These two cone-like bands repel each other when they cross.
For $\zeta=ja/2$ with $j$ being an integer, in particular, we have a bilayer with AA stacking and $v_1=\gamma_1$, $v_2=0$, and $e^{i\varphi_1}=\omega^{-j-1}$.
In this case, the two cone-like bands have different energies by $\pm\gamma_1$ and do not interact each other.
For $\zeta=ja/4$, we have $\varphi_2=\varphi_1$, for which two cone-like bands displaced from each other both in wave-vector and energy become independent.
No AB stacked bilayer is formed.
\par
%
Figure \ref{Fig:Structure_of_Flattened_Armchair_Nanotube} shows some examples of the structure of the flattened region for armchair tubes ($\eta=\pi/6$).
Inter-valley components identically vanish, because the K and K' points are mapped onto themselves after folding.
An A site, however, turns into a B site and a B site turns into an A site, respectively.
The effective potential is again independent of the position well inside the flattened region, but varies periodically as a function of $\zeta$ with period $3b/2$, where $b=a/\sqrt3$.
We can set $\zeta = b ( 3 j + p )/2$, with integer $j$ and $0\!\le\!p\!<\!3$.
\par
%
Numerical calculations show
%
\begin{eqnarray}
\tilde V_{AA} \!\! & = & \!\! - v_{A} \pmatrix{1 & 0 \cr 0 & 1 \cr } , \quad
\tilde V_{BB} = - v_{B} \pmatrix{1 & 0 \cr 0 & 1 \cr } \nonumber \\
\noalign{\vspace{-0.250cm}}
\\
\noalign{\vspace{-0.250cm}}
\tilde V_{AB} \!\! & = & \!\! \tilde V_{BA}^\dagger = i \, v_{AB} \pmatrix{1 & 0 \cr 0 & 1 \cr } , \nonumber
\end{eqnarray}
%
with real $v_{A}$, $v_{B}$, and $v_{AB}$ varying as a function of $\zeta$.
For $p=0$ and 1, we have a bilayer with AB stacking and for $p=2$, we have a bilayer with AA stacking.
In fact, for $p=0$ we have $v_A=v_{AB}=0$ and $v_B=\gamma_1$, for $p=1$ we have $v_A=\gamma_1$ and $v_B=v_{AB}=0$, and for $p=2$ we have $v_A=v_B=0$ and $v_{AB}=\!-\!\gamma_1$.
\par
%
In the following, we shall consider effects of inter-wall interactions in the case of narrow flattened region $L_F/L\!\ll\!1$ by perturbation analysis.
This analysis is useful for understanding qualitative features of inter-wall interactions appearing in numerically obtained band structure as shown in the next section.
\par
%
Because the effective inter-wall potential is independent of position, the spatial part of the matrix element is given by an overlapping integral.
For KK elements, for example, we have
%
\begin{eqnarray}
S_{nn'}^{KK} \!\! & = & \!\! {1 \over L} \bigg[ \int_{(L-L_F)/4}^{(L+L_F)/4} \! + \! \int_{(-L-L_F)/4}^{(-L+L_F)/4} \bigg] e^{ i [\kappa_\nu(n)+\kappa_\nu(n')] x} d x \nonumber \\
& = & \!\! {2\over \pi} {1 \over n + n' - {2\over 3} \nu } \sin\Big[{\pi \over 2} {L_F \over L} \Big( n + n' - {2\over 3} \nu \Big) \Big] \nonumber \\
\noalign{\vspace{-0.150cm}}
&& \!\! \times \cos\Big[{\pi \over 2}\Big(n + n' - {2\over 3}\nu \Big) \Big] ,
\end{eqnarray}
%
which for $L_F/L\!\ll\!1$ becomes
%
\begin{equation}
S_{nn'}^{KK} = {L_F \over L} \cos\Big[{\pi \over 2}\Big(n + n' - {2\over 3}\nu \Big) \Big] .
\end{equation}
%
The matrix element $S_{nn'}^{K'K'}$ can be obtained from $S_{nn'}^{KK}$ by replacing $\nu$ with $-\nu$.
Further, $S_{nn'}^{KK'}$ and $S_{nn'}^{K'K}$ for different valleys can be obtained by setting $\nu=0$ in the above.
\par
%
\subsection{Weak Inter-Wall Coupling: Zigzag Tube} \label{Ssc:Weak_Inter-Wall_Coupling:_Zigzag_Tube}
%
For zigzag tubes, the effective inter-wall potential causes coupling between the K and K' points.
In the following, we shall consider the case that the flattened region has the structure of an AA-stacked bilayer, i.e., $\zeta=j(a/2)$ with integer $j$.
\par
%
In the case of semiconducting tubes ($\nu=\pm1$), degenerate states associated with the K and K' points are characterized by $s'=s$ and $|\kappa_\nu(n)|=|\kappa_{-\nu}(n')|$, giving $n'=-n$ or $\kappa_{-\nu}(n')=-\kappa_\nu(n)$, as shown in Fig.\ \ref{Fig:Nanotube_Bands} (c).
The matrix elements are calculated as
%
\begin{equation}
\big[V^{KK'}\big]_{ns,-ns} = \big[V^{K'K\dagger}\big]_{ns,-ns} = \omega^{-j-1} {\gamma_1 L_F \over L} ,
\label{Eq:Perturbation_Zigzag_Semiconducting}
\end{equation}
%
independent of $n$.
This shows that the two degenerate states split into two by the amount $\pm\gamma_1 L_F/L$ independent of bands.
As will be shown in the next section, this can convert the tube into metallic for sufficiently large $L_F/L$.
\par
%
In metallic case $\nu=0$, there are two degenerate metallic linear bands for $n=n'=0$ as shown in Fig.\ \ref{Fig:Nanotube_Bands} (b).
The matrix elements become
%
\begin{equation}
\big[V^{KK'}\big]_{0s,0s'} = \big[V^{K'K\dagger}\big]_{0s,0s'} = \omega^{-j-1} (1 + ss') {\gamma_1 L_F \over 2 L} .
\label{Eq:Perturbation_Zigzag_Metallic_Linear}
\end{equation}
%
This shows that two degenerates states associated with the K and K' points split into two by the amount $\pm\gamma_1 L_F/(2L)$ independent of bands and there is no band-gap opening.
\par
%
For parabolic bands $n=\pm n_0$ and $n'=\pm n_0$ with $n_0\!>\!0$ and $s'=s$, the matrix elements are calculated up to linear order in $k$, and the effective Hamiltonian within the four degenerate states becomes
%
\begin{equation}
{\cal H}_{\rm eff} = - {\gamma_1 L_F \over L} \!\!\! 
\bordermatrix{ & +n_0 K & -n_0 K & +n_0 K' & -n_0 K' \cr
& 0 & 0 & \!\! i \omega^{-j-1} \delta \!\! & - \omega^{-j-1} \cr
& 0 & 0 & \!\! - \omega^{-j-1} & \!\! -i \omega^{-j-1} \delta \! \cr
& \! -i \omega^{j+1} \delta & - \omega^{j+1} \!\! & 0 & 0 \cr
& - \omega^{j+1} & \!\! i \omega^{j+1} \delta \!\! & 0 & 0 \cr } ,
\end{equation}
%
with $\delta \approx (-1)^{n_0} k /\kappa_0(n_0)$.
With the use of the unitary matrix
%
\begin{equation}
U = {1\over \sqrt2} \pmatrix{ 1 & 1 & 0 & 0 \cr
0 & 0 & 1 & 1 \cr
0 & 0 & - \omega^{j+1} & \omega^{j+1} \cr
- \omega^{j+1} & \omega^{j+1} & 0 & 0 \cr } ,
\end{equation}
%
the effective Hamiltonian is converted into
%
\begin{equation}
U^\dagger {\cal H}_{\rm eff} U = - \gamma_1 {L_F \over L} \pmatrix{ +1 & 0 & - i \delta & 0 \cr
0 & -1 & 0 & +i \delta \cr
+i \delta & 0 & +1 & 0 \cr
0 & -i \delta & 0 & - 1 \cr } .
\label{Eq:Perturbation_Zigzag_Metallic_Parabolic}
\end{equation}
%
This shows that the two bands, each doubly degenerate, are split by the amount $\pm\gamma_1 L_F/L$ and then the remaining degeneracy is further lifted by $\delta$ which is proportional to the wave vector and inversely to $|n|$.
\par
%
\begin{figure*}
\begin{center}
\includegraphics[width=7.0cm]{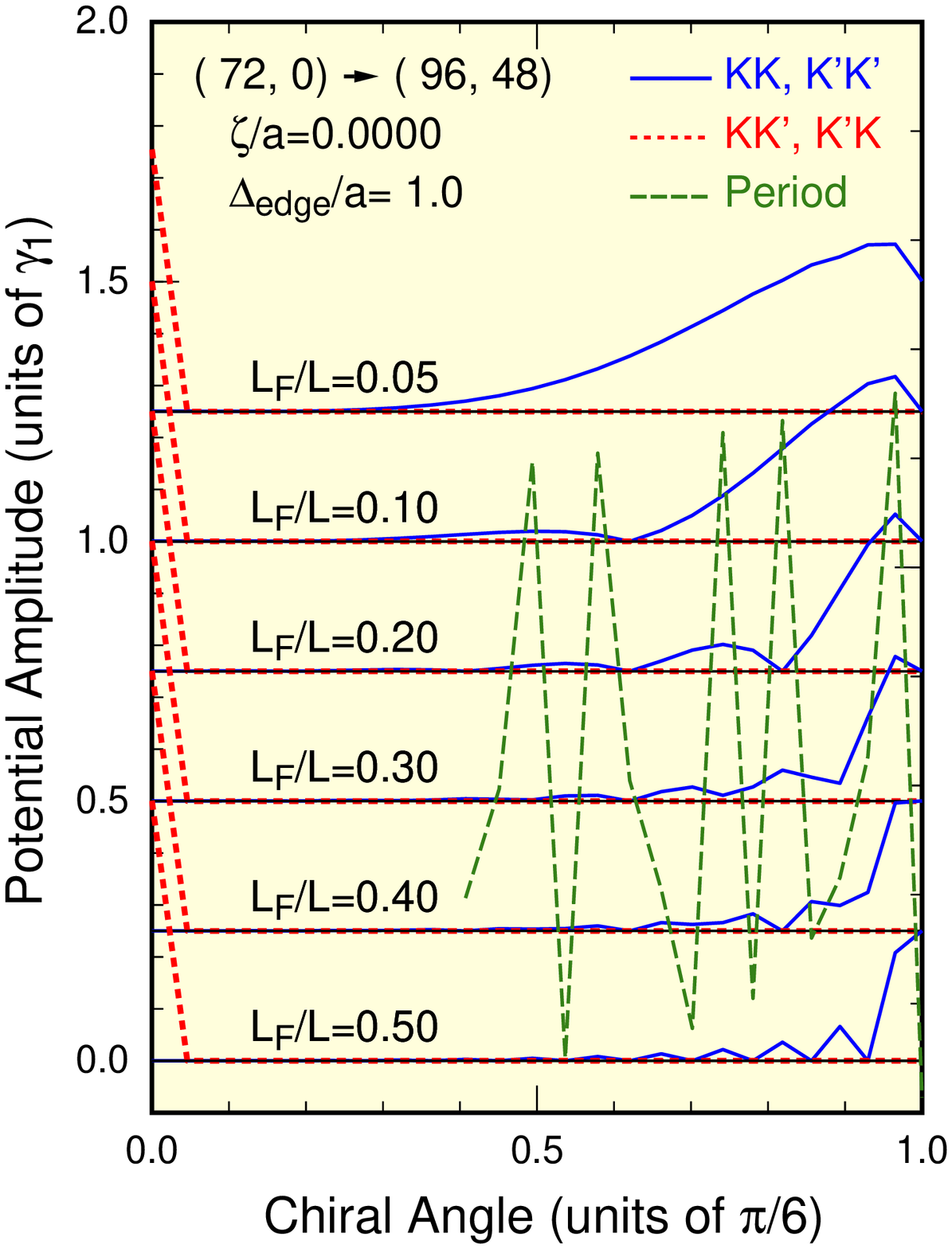}
\hskip-1.80cm
\includegraphics[width=7.0cm]{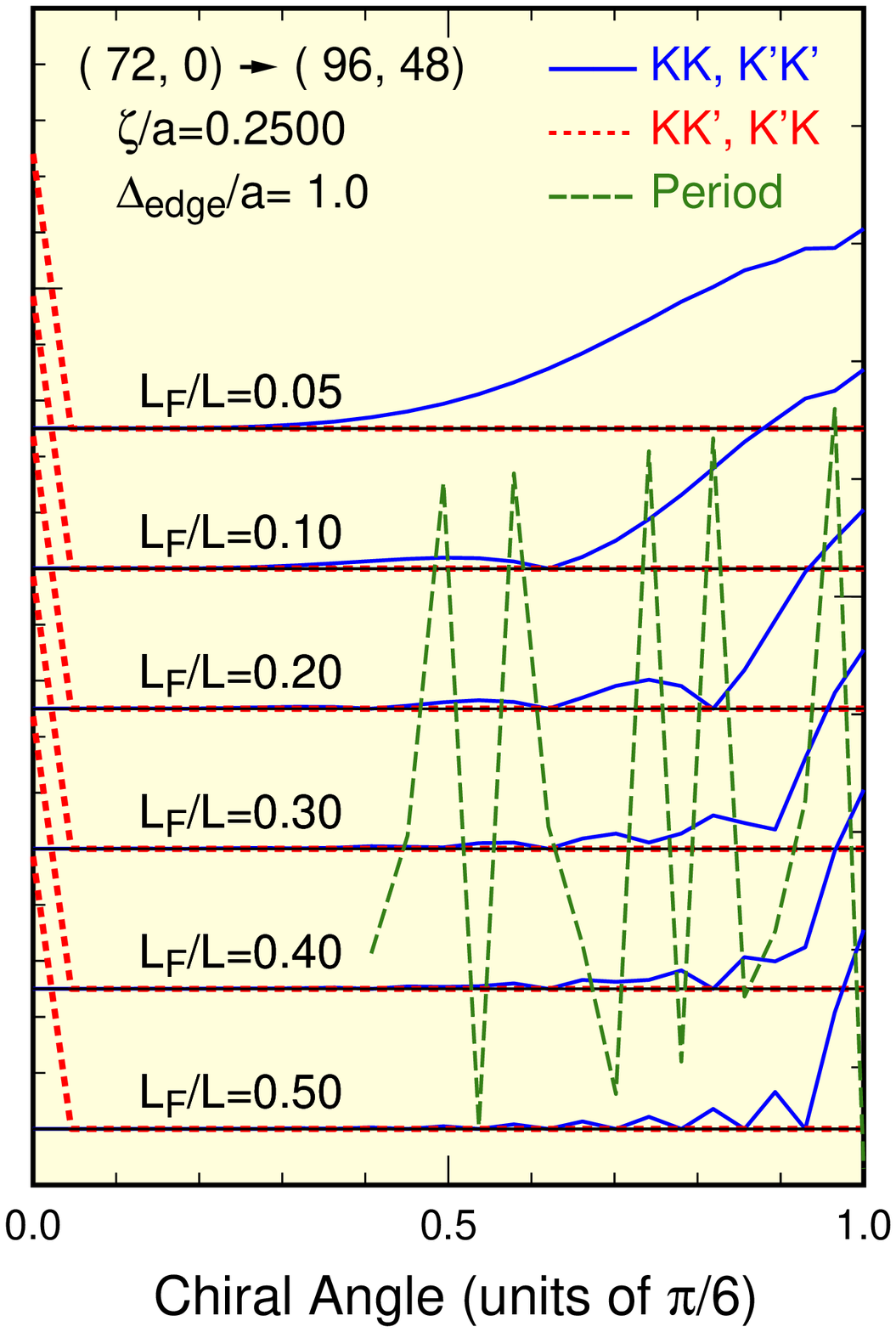}
\hskip-1.80cm
\includegraphics[width=7.0cm]{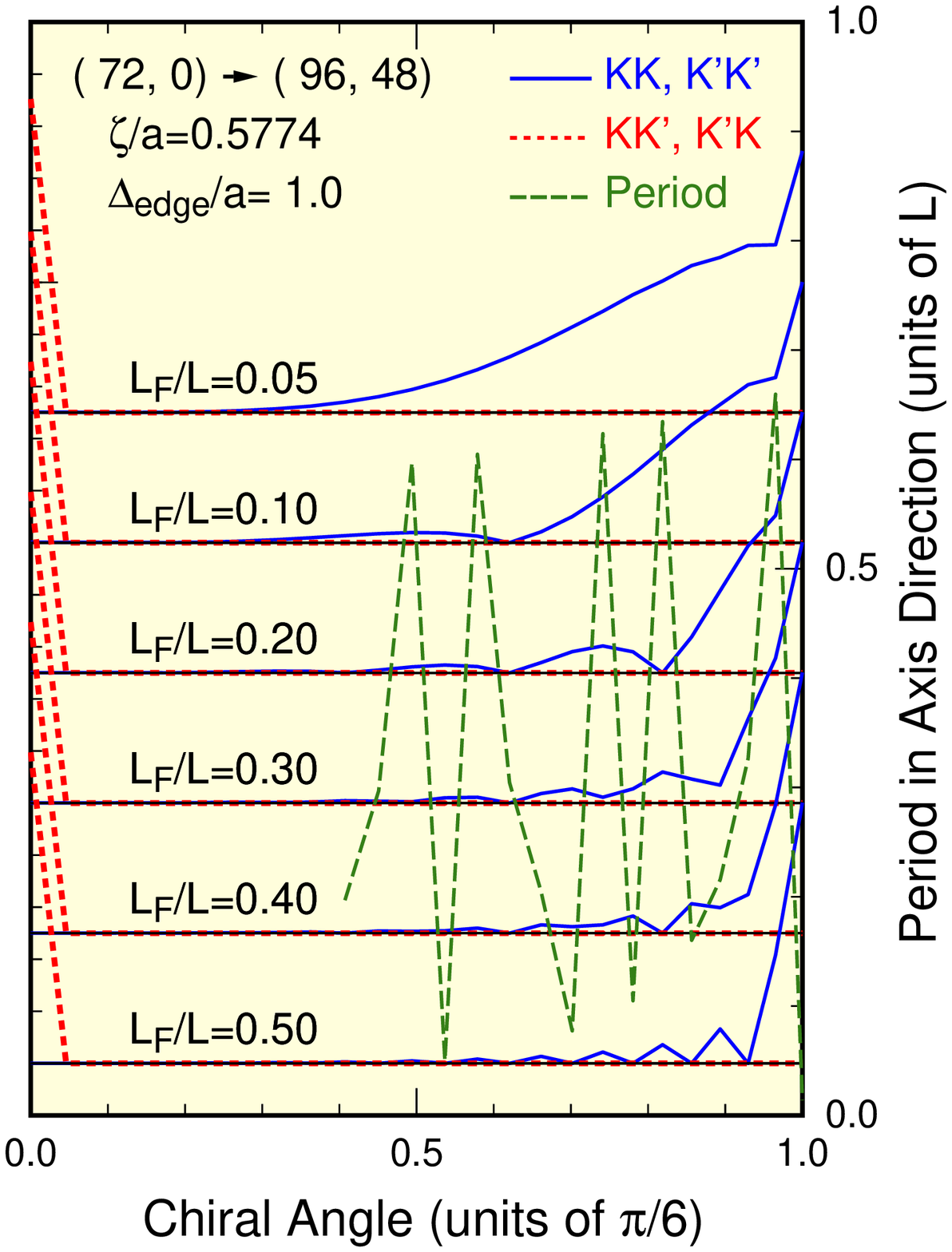}
\caption{%
\label{Fig:Dominant-Term_Approximation:_1}
Some examples of the potential amplitude in the dominant-term approximation as the function of chiral angle $\eta$ with family index $f=144$ defined in Eq.\ (\ref{Eq:Family_Number:_Definition}), i.e., $(n_a,n_b)=(72,0)$, (73,2), $\dots$, (96,48).
The average of the absolute values of each element of $\hat V$ are shown for $\hat V^{KK}$ and $\hat V^{K'K'}$ by (blue) solid lines and for $\hat V^{KK'}$ and $\hat V^{K'K}$ by (red) dotted lines.
The size of the one-dimensional unit cell $T$ is also shown in units of $L$ by (green) dashed lines.
We chose relative displacement $\zeta/a=0$, 1/4, and $1/\sqrt3=0.5774\cdots$ defined in Fig.\ \ref{Fig:Flattened_CN} for three panels and a parameter $\Delta_{\rm edge}$ is defined in Eq. (\ref{Eq:Delta_edge}).
}
\end{center}
\vspace{-0.50cm}
\end{figure*}
%
\subsection{Weak Inter-Wall Coupling: Armchair Tube} \label{Ssc:Weak_Inter-Wall_Coupling:_Armchair_Tube}
%
Armchair nanotubes have $\nu=0$ and $\mu=\pm1$, and therefore are always metallic in the absence of inter-wall interaction, as shown in Fig.\ \ref{Fig:Nanotube_Bands} (a).
Thus, dominant inter-wall coupling is present only within each of the K and K' points and $S_{nn'}^{KK}=S_{nn'}^{K'K'}$.
For $p=0$ with AB-stacking structure, the matrix elements are calculated as
%
\begin{equation}
\big[ V^{KK} \big]_{ns,n's'} = \big[ V^{K'K'} \big]_{ns,n's'} = - {s s' \over 2} \gamma_1 S_{nn'}^{KK} .
\end{equation}
%
For $n=n'=0$, there are two degenerate states $s=\pm 1$ at $k=k_\mu$ corresponding to the K point, and the effective Hamiltonian becomes
%
\begin{equation}
{\cal H}_{\rm eff} = \pmatrix{ \displaystyle + \gamma|k-k_\mu| - {\gamma_1 L_F \over 2 L} \!\!\!\! & \displaystyle + {\gamma_1 L_F \over 2L} \!\!\!\! \cr \displaystyle \!\!\!\! + { \gamma_1 L_F \over 2 L} & \displaystyle \!\!\!\! - \gamma|k-k_\mu| - { \gamma_1 L_F \over 2 L} \cr } ,
\label{Eq:Perturbation_Armchair_AB_Linear}
\end{equation}
%
which gives the bands
%
\begin{equation}
\varepsilon = - {\gamma_1\over 2} {L_F \over L}\pm \sqrt{\gamma^2 (k-k_\mu)^2 + \Big({\gamma_1 L_F \over 2L} \Big)^2 } .
\end{equation}
%
This shows that the bottom of the conduction band remains at zero energy while the top of the valence band is lowered by $\gamma_1L_F/L$.
\par
%
Because parabolic bands $\pm n_0$ with $n_0>0$ and $s'=s$ are degenerate, the effective Hamiltonian becomes
%
\begin{equation}
{\cal H}_{\rm eff} = - { \gamma_1 L_F \over 2 L} \pmatrix{ (-1)^{n_0} & 1 \cr 1 & (-1)^{n_0} \cr } ,
\label{Eq:Perturbation_Armchair_AB_Parabolic}
\end{equation}
%
giving energy shift of $-[(-1)^{n_0} \pm 1] \gamma_1 L_F/(2L)$.
This results in an alternate upward or downward shift by $(-1)^{n_0+1}\gamma_1L_F/L$ for a band and no shift for another.
For $p=1$ with another AB-stacking structure, exactly the same results can be obtained by changing the phase of the wave functions in an appropriate manner.
\par
%
For the case of AA-stacking ($p=2$) shown in Fig.\ \ref{Fig:Structure_of_Flattened_Armchair_Nanotube} (d), the matrix element for the K point becomes
%
\begin{eqnarray}
\big[ V^{KK} \big]_{ns,n's'} \!\! & = & \!\! - { i \over 2} \gamma_1 \Big( {s' [ \kappa_0(n) + i (k-k_\mu)] \over \sqrt{\kappa_0(n)^2 + (k-k_\mu)^2 } } \nonumber\\
&& \!\! - {s [ \kappa_0(n') - i (k-k_\mu)] \over \sqrt{\kappa_0(n')^2 + (k-k_\mu)^2 } } \Big) S_{nn'}^{KK} .
\end{eqnarray}
%
For $n'=-n$ including $n=0$, we have $\kappa_0(n')=-\kappa_0(n)$ and therefore,
%
\begin{equation}
\big[ V^{KK} \big]_{ns,-ns'} = - { i \over 2} (s + s') { \kappa_0(n) + i (k-k_\mu) \over \sqrt{\kappa_0(n)^2 + (k-k_\mu)^2 } } {\gamma_1 L_F \over L}.
\label{Eq:Effective_Hamiltonian:_Armchair_AA_Stacking}
\end{equation}
%
For the metallic linear bands ($n=0$), there are two degenerate states $s=\pm1$.
Off diagonal elements $s'=-s$ in Eq. (\ref{Eq:Effective_Hamiltonian:_Armchair_AA_Stacking}) vanish, showing that there is no splitting.
The bands are shifted by diagonal elements
%
\begin{equation}
\big[ V^{KK} \big]_{0s,0s} = + {s (k-k_\mu) \over |k-k_\mu|} { \gamma_1 L_F \over L}.
\label{Eq:Perturbation_Armchair_AA_Linear}
\end{equation}
%
This corresponds to a parallel shift in the negative $k$ direction.
In contrast, the linear bands at the K' point shift in the positive $k$ direction.
\par
%
For parabolic bands $s'=s$, off-diagonal elements $n'\allowbreak=-n\ne0$ in Eq.\ (\ref{Eq:Effective_Hamiltonian:_Armchair_AA_Stacking}) causes the splitting of $\pm\gamma_1 L_F/L$ independent of the bands.
For small $k$, diagonal elements for $n'=n \ne 0$ become
%
\begin{eqnarray}
\big[ V^{KK} \big]_{ns,ns} \!\! & \approx & \!\! (-1)^n { s (k-k_\mu) \over |\kappa_0(n)| } {\gamma_1 L_F \over L} ,
\label{Eq:Perturbation_Armchair_AA_Parabolic}
\end{eqnarray}
%
corresponding to a parallel shift in $k$ direction with amount decreasing with $|n|$ as $|n|^{-1}$.
Sign of the parallel shift is positive for odd $n$ and negative for even $n$.
\par
%
\subsection{Chiral Tubes: Dominant Terms}
%
It is known that essential properties of carbon nanotubes can be specified by family index $f$ given by
%
\begin{equation}
f = 2 n_a - n_b .
\label{Eq:Family_Number:_Definition}
\end{equation}
%
For ${\bf L}$ having $f$, we have ${\bf L} \cdot {\bf a} = {1\over 2} f a^2$, which means that the chiral vector lies on the line vertical to the horizontal axis with distance $(1/2)fa$ from the origin.
From Eq.\ (\ref{Eq:Family_Number:_Definition}), we have $n_a + n_b = 3 n_a - f$, meaning that the value of $\nu$ is determined by $f$.
Therefore, tubes with a given value of $f$ have the same value of $\nu$ and always metallic or semiconducting independent of individual values of $n_a$ and $n_b$.
\par
%
The dominant contribution of effects of inter-wall coupling in the flattened region may be estimated from the Fourier coefficients of small $n$ and $m$.\cite{Mele_2010a,Lopes_dos_Santos_et_al_2012a}
It is natural to choose $n=0$ for both intra- and inter-valley terms.
For intra-valley terms we choose $m=0$ and for inter-valley terms we choose $m=+\mu$ for $V^{KK'}$ and $m=-\mu$ for $V^{K'K}$.
\par
%
In the following, in order to show the magnitude of the effective potential, we plot $(L/L_F)V_{n,m}^{KK}$, etc.\ instead of $V_{n,m}^{KK}$, etc.\ themselves.
Figure \ref{Fig:Dominant-Term_Approximation:_1} shows the average of the absolute value of each element $\hat V$ separately for intra-valley (KK and K'K') and inter-valley elements (KK' and K'K) as a function of the chiral angle $\eta$ for tubes with family number $f=144$, i.e., $(n_a,n_b)=(72,0)$, (73,2), $\cdots$, (96,48).
These tubes have $\nu=0$ and therefore are metallic.
We have chosen the cases of $\zeta/a=0$, 1/4, and $1/\sqrt3=0.5774\cdots$.
\par
%
\begin{figure*}
\begin{center}
\begin{minipage}[t]{16.0cm}
(a) \hskip-0.750cm \includegraphics[height=7.50cm]{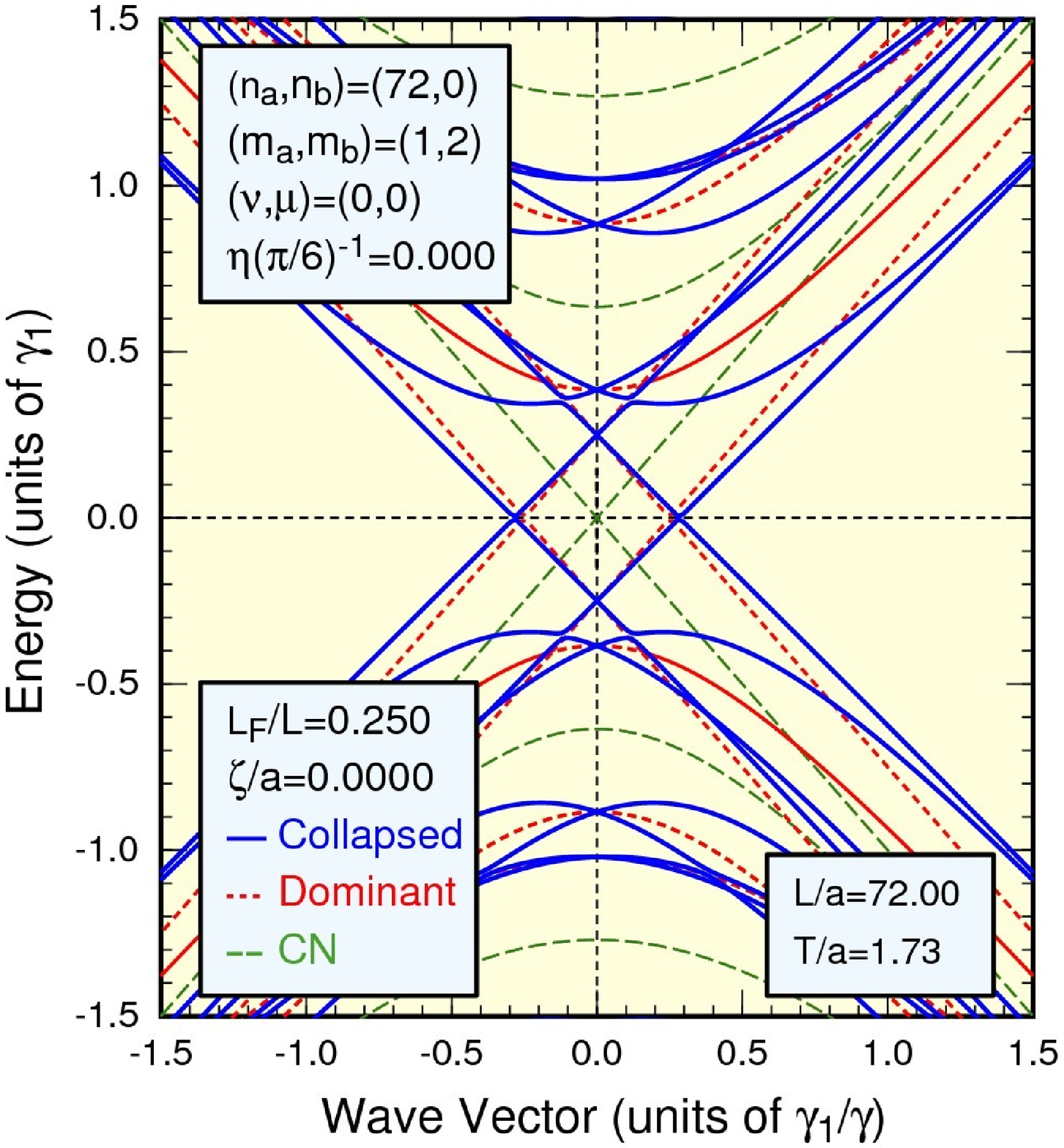}
\enspace
(b) \hskip-0.750cm \includegraphics[height=7.50cm]{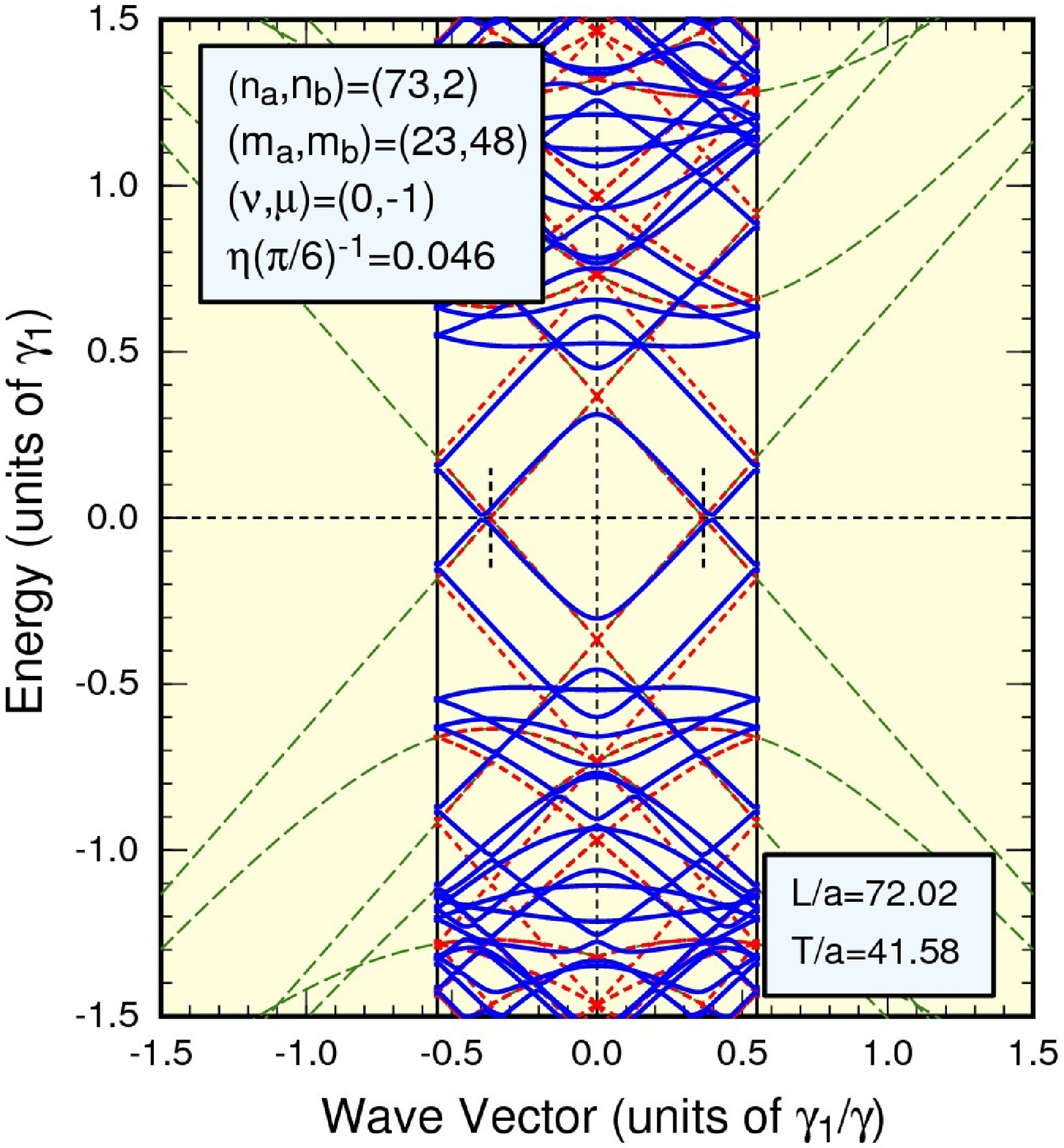}
\end{minipage}
\begin{minipage}[t]{16.0cm}
(c) \hskip-0.750cm \includegraphics[height=7.50cm]{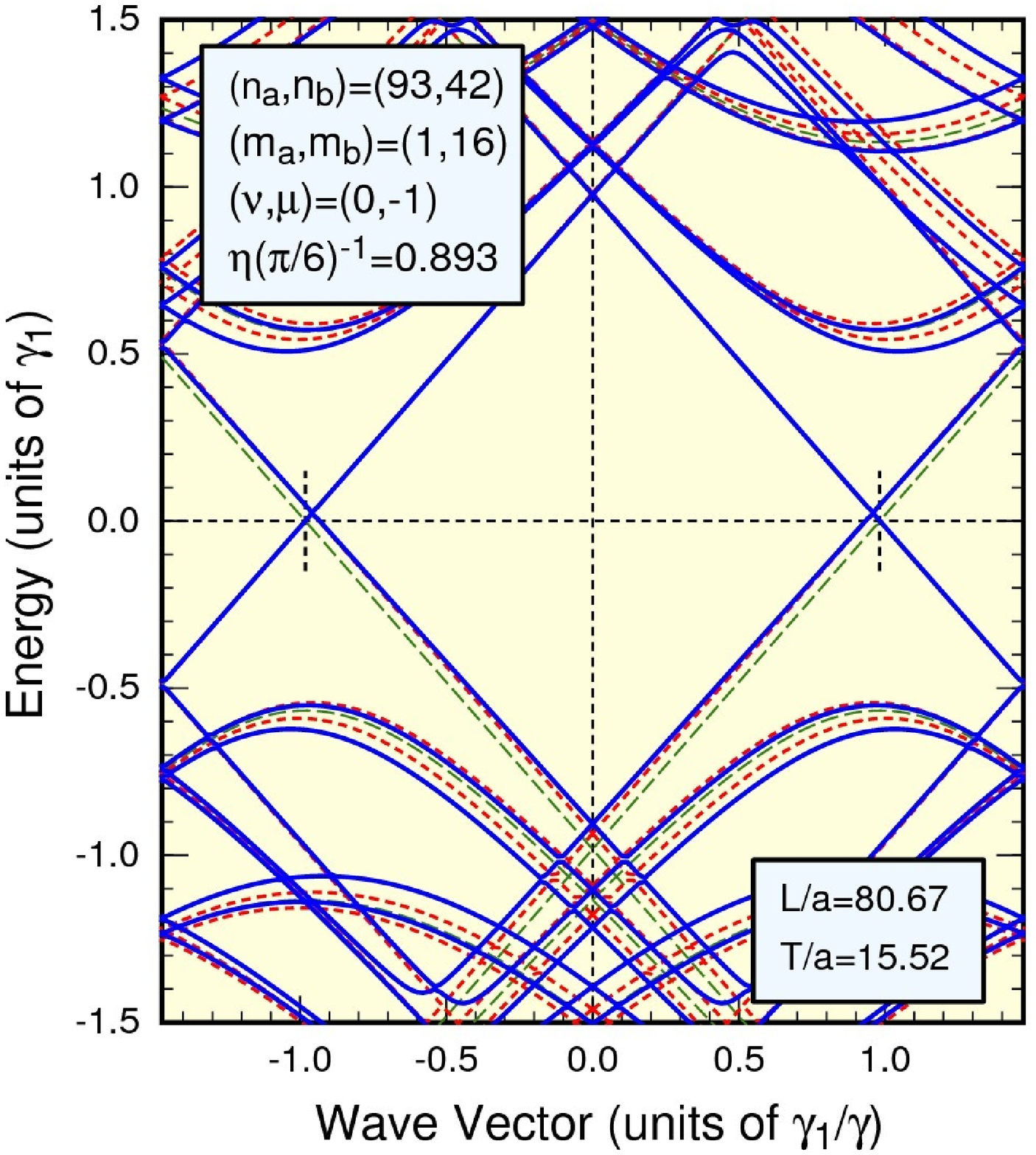}
\enspace ~
(d) \hskip-0.750cm \includegraphics[height=7.50cm]{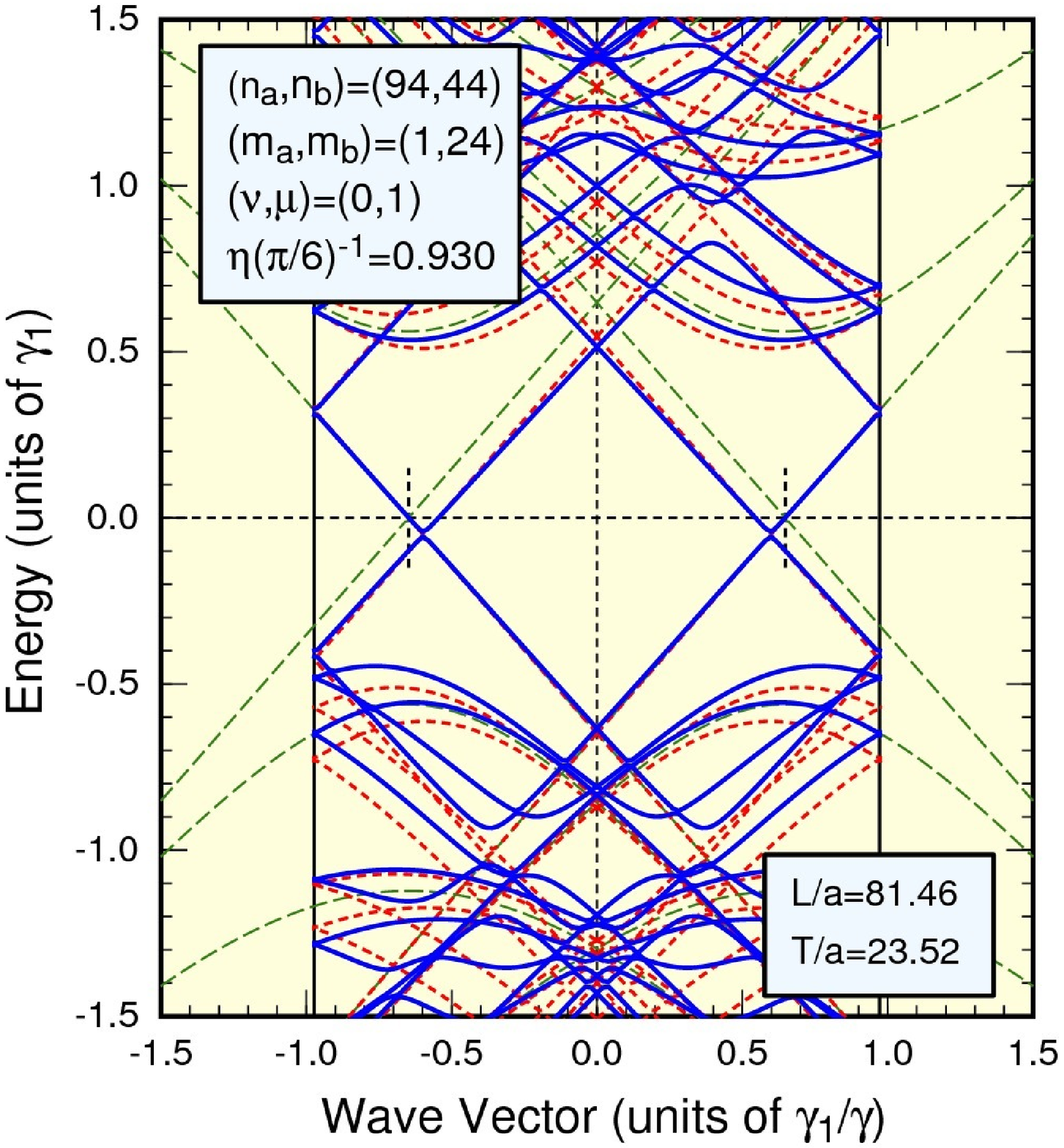}
\end{minipage}
\begin{minipage}[t]{16.0cm}
(e) \hskip-0.750cm \includegraphics[height=7.50cm]{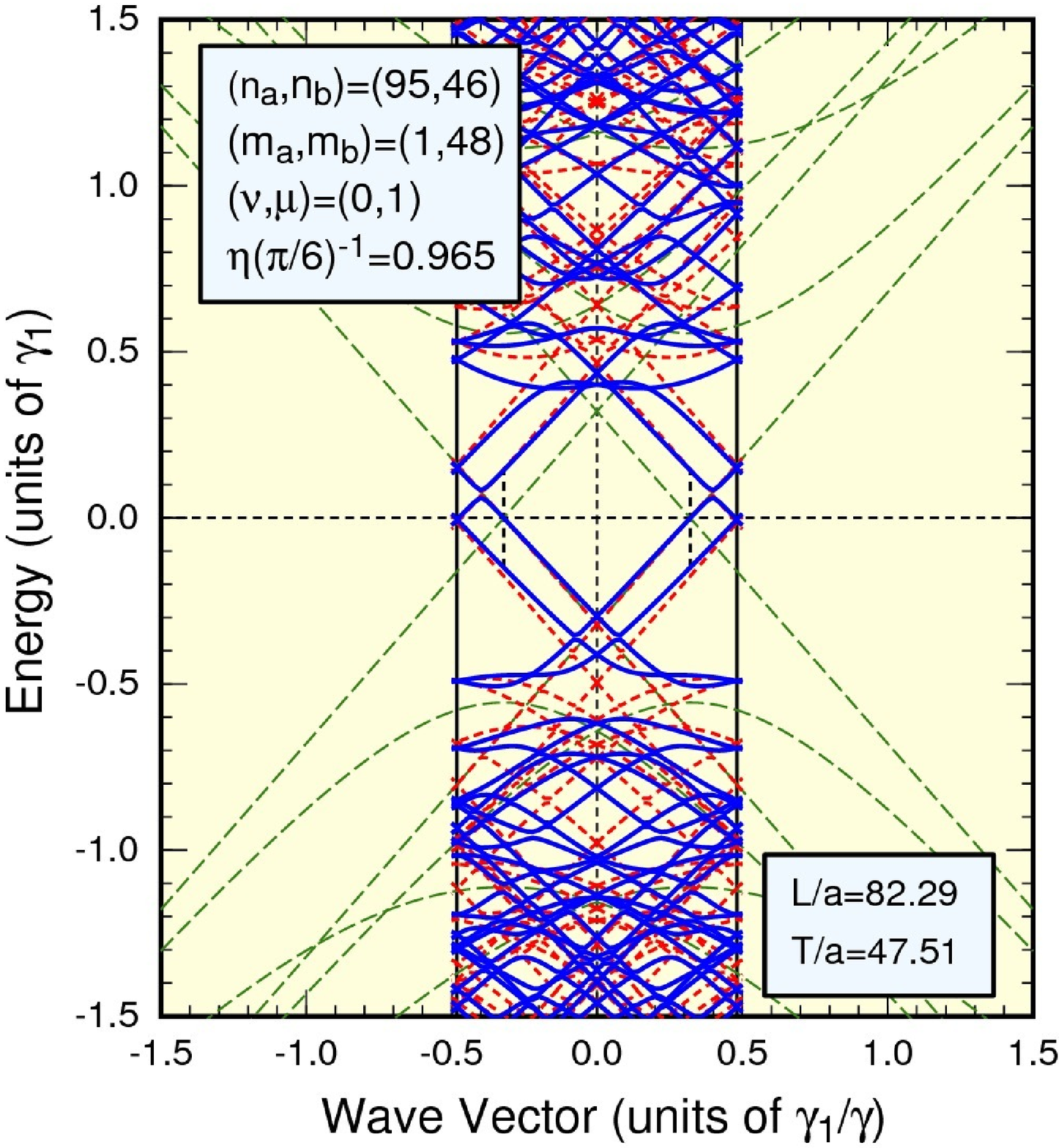}
\enspace
(f) \hskip-0.750cm \includegraphics[height=7.50cm]{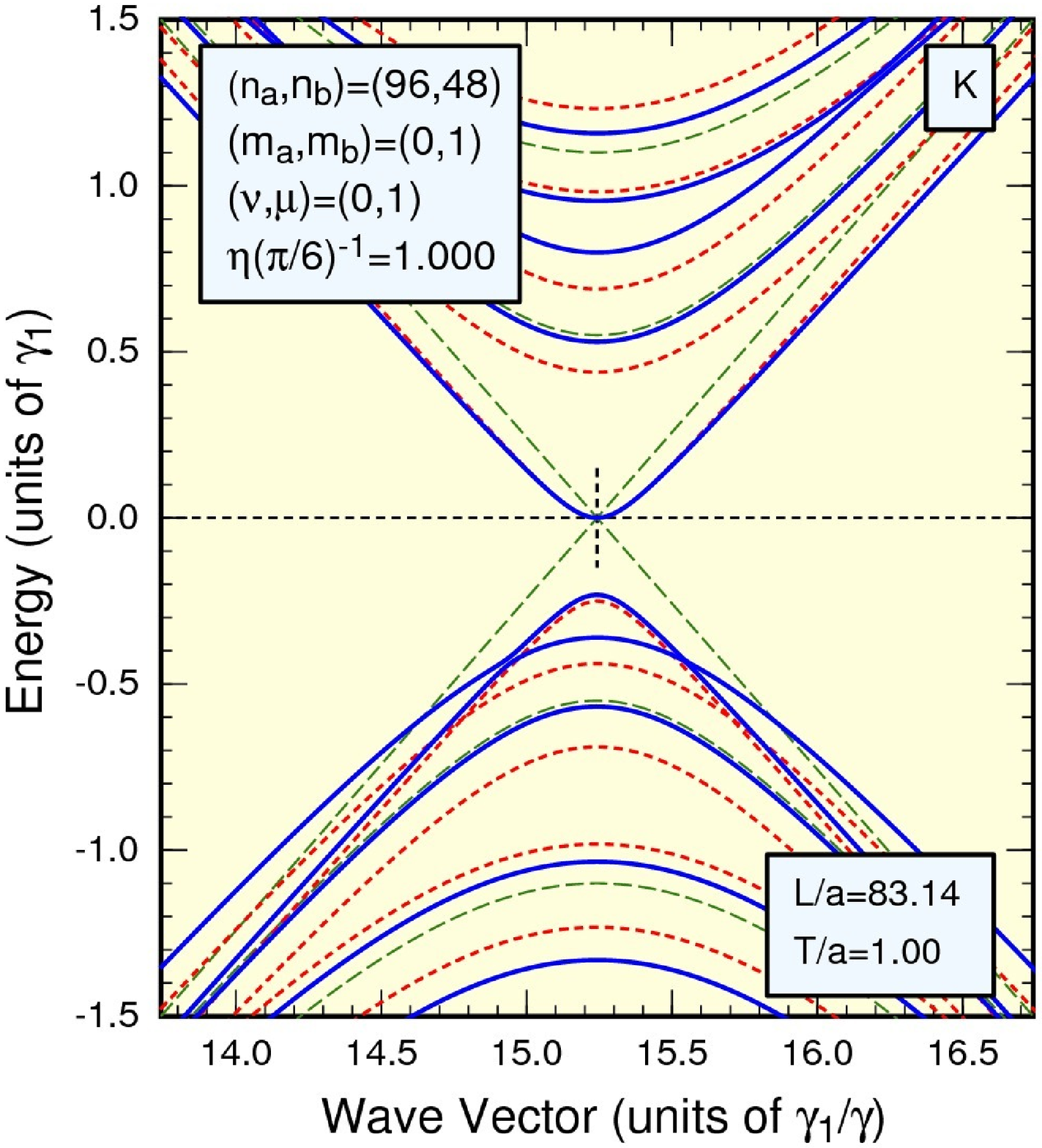}
\end{minipage}
\caption{%
\label{Fig:Band_Structure_(Metallic_zeta=0)}
Calculated band structure of collapsed tubes with family index $f=144$ having zigzag (a), its neighboring structures (b), armchair (f), and its neighboring structures (e) and (d).
Parameters are defined in Eqs.\ (\ref{Eq:nu}), (\ref{Eq:TranslationVector}), and (\ref{Eq:mu}), and illustrated in Fig.\ \ref{Fig:Flattened_CN}.
}
\end{center}
\vspace{-0.50cm}
\end{figure*}
%
\begin{figure*}
\begin{center}
\begin{minipage}[t]{16.0cm}
(a) \hskip-0.750cm \includegraphics[height=7.50cm]{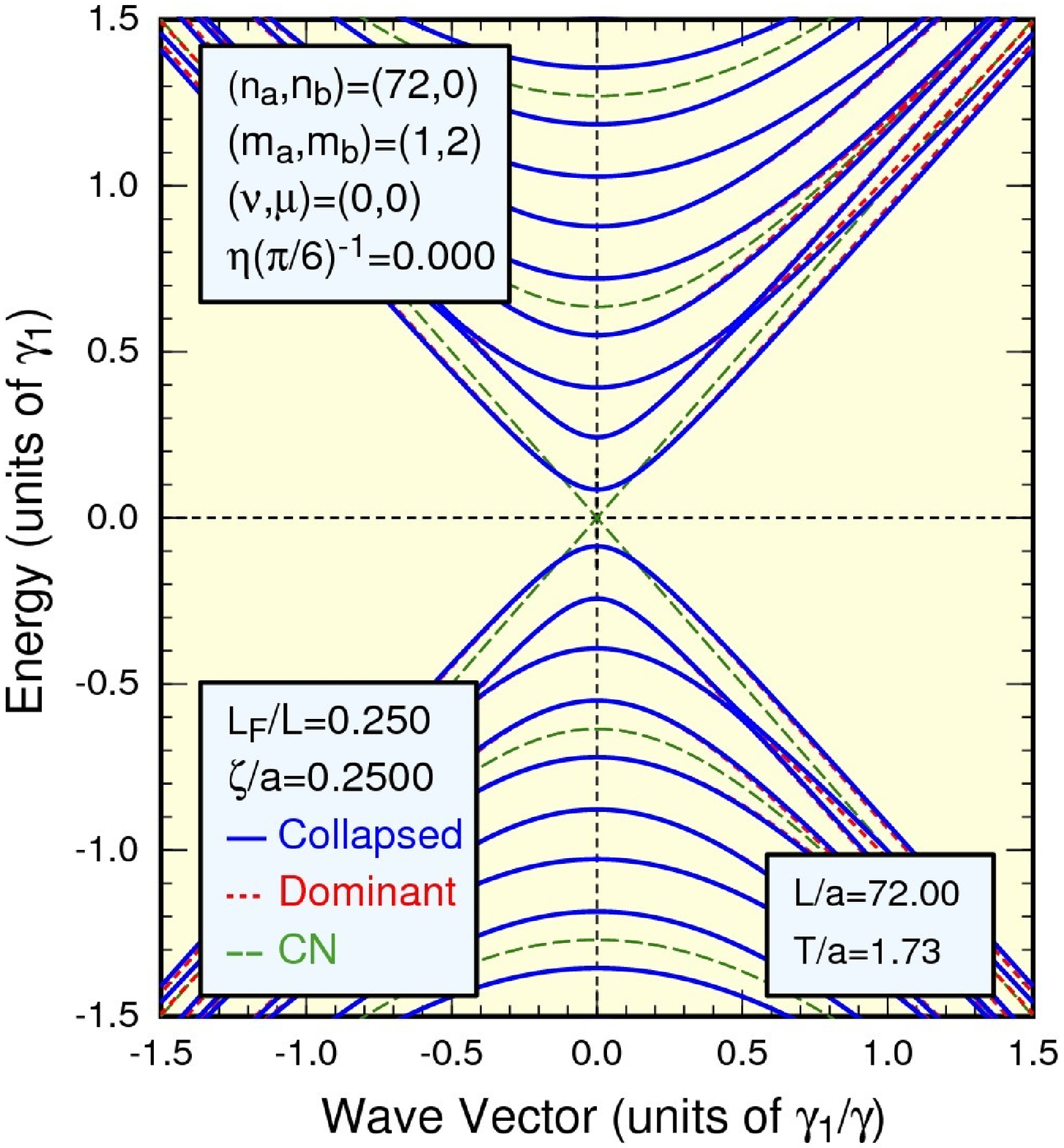}
\enspace
(b) \hskip-0.750cm \includegraphics[height=7.50cm]{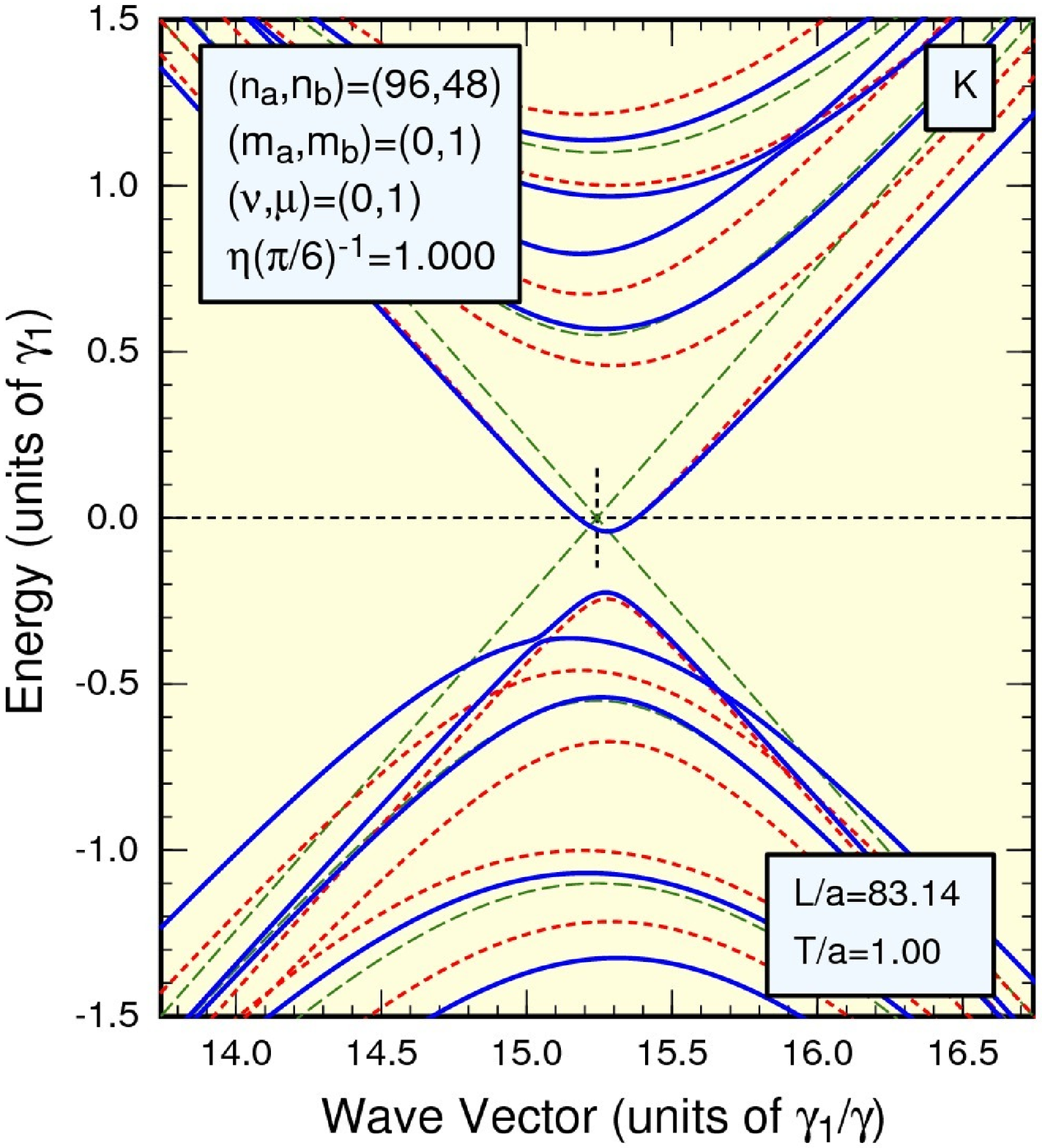}
\end{minipage}
\begin{minipage}[t]{16.0cm}
(c) \hskip-0.750cm \includegraphics[height=7.50cm]{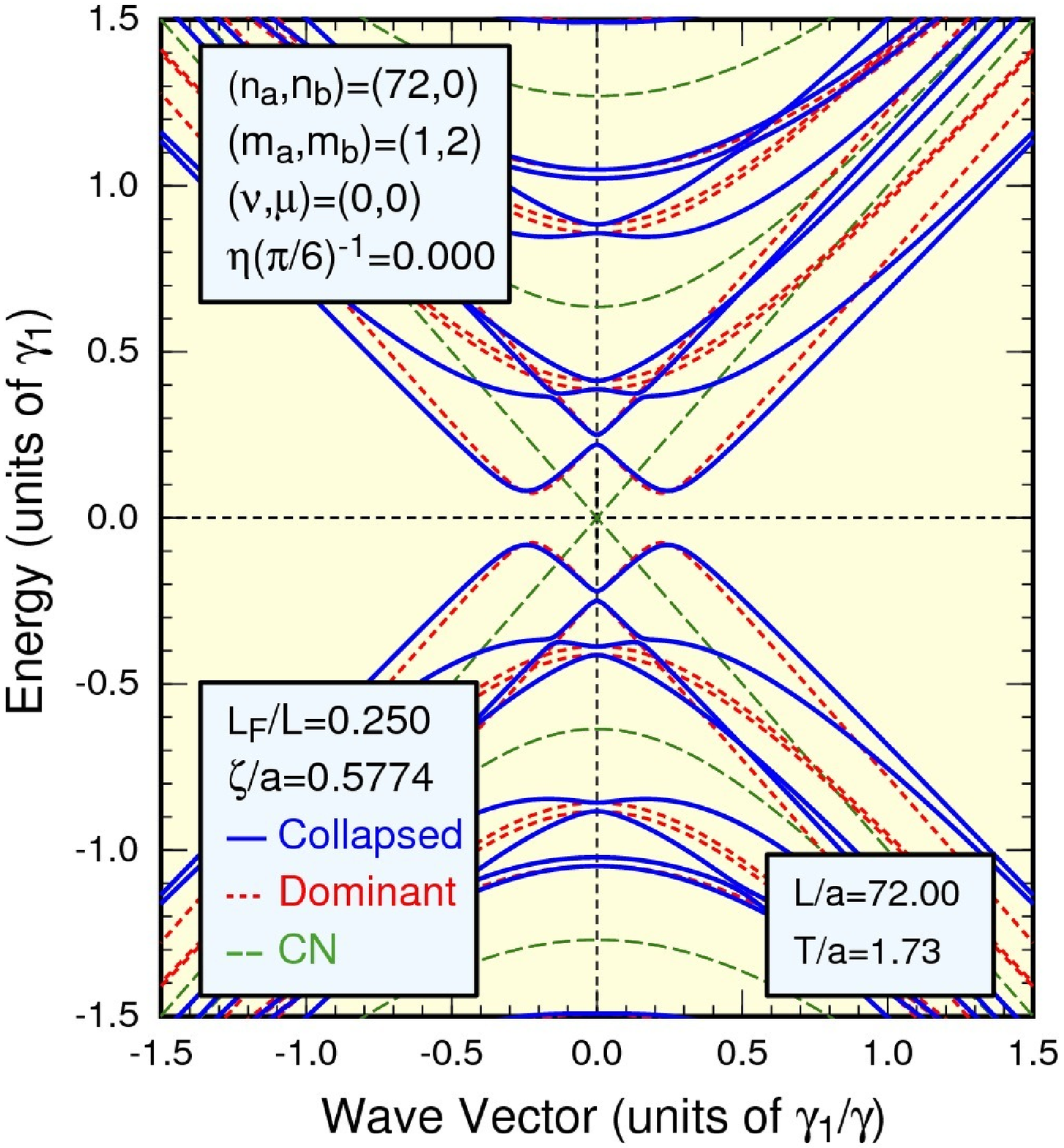}
\enspace
(d) \hskip-0.750cm \includegraphics[height=7.50cm]{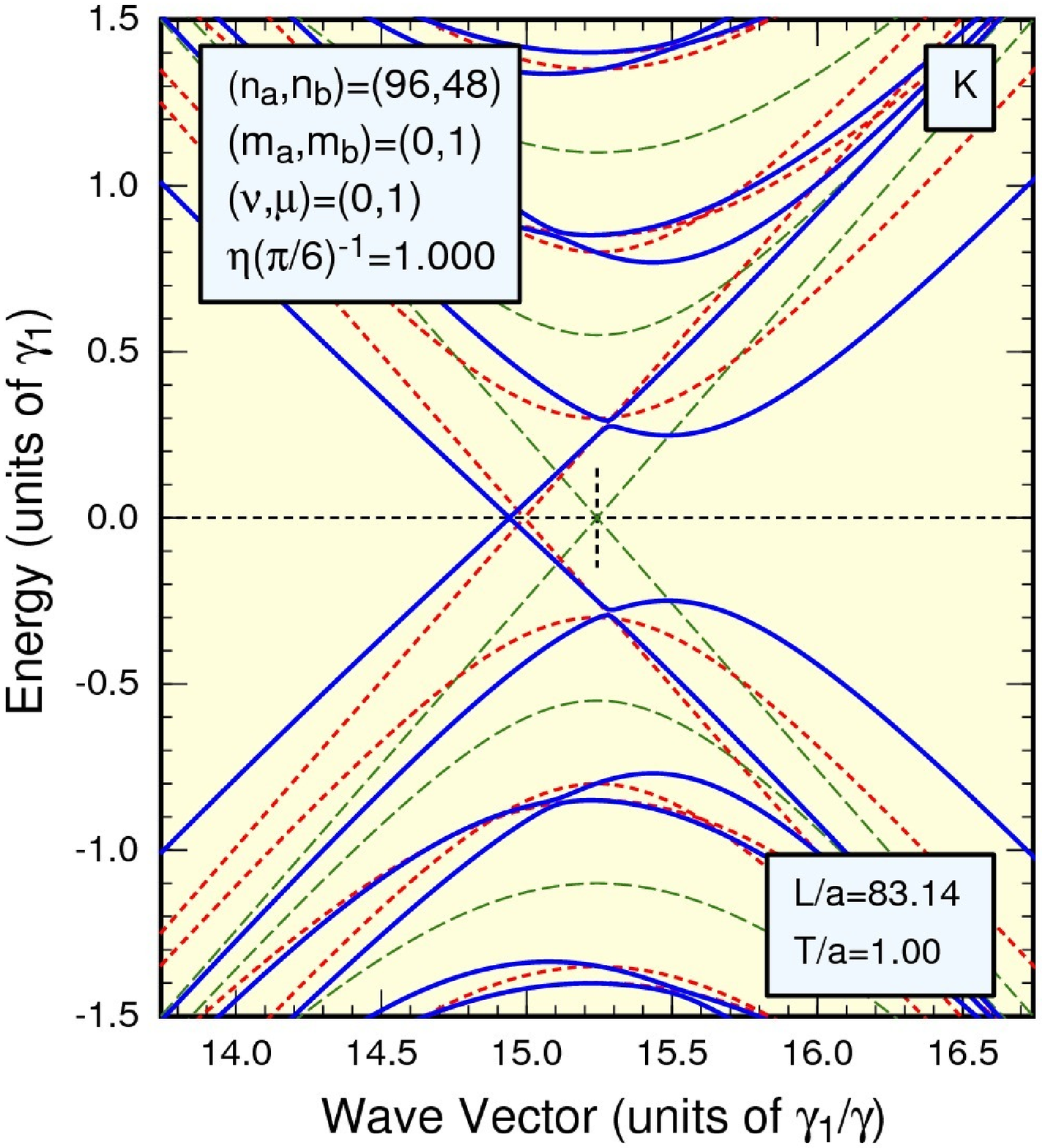}
\end{minipage}
\caption{%
\label{Fig:Band_Structure_(Metallic,_Zigzag,_zeta=1/4,_1/sqrt3)}
Calculated band structure of collapsed tubes with $f=144$ having zigzag and its neighboring structures.
(a) and (b) $\zeta/a=1/4$.
(c) and (d) $\zeta/a=1/\sqrt3$.
}
\end{center}
\vspace{-0.50cm}
\end{figure*}
%
\begin{figure*}
\begin{center}
\begin{minipage}[t]{16.0cm}
(a) \hskip-0.750cm \includegraphics[height=7.50cm]{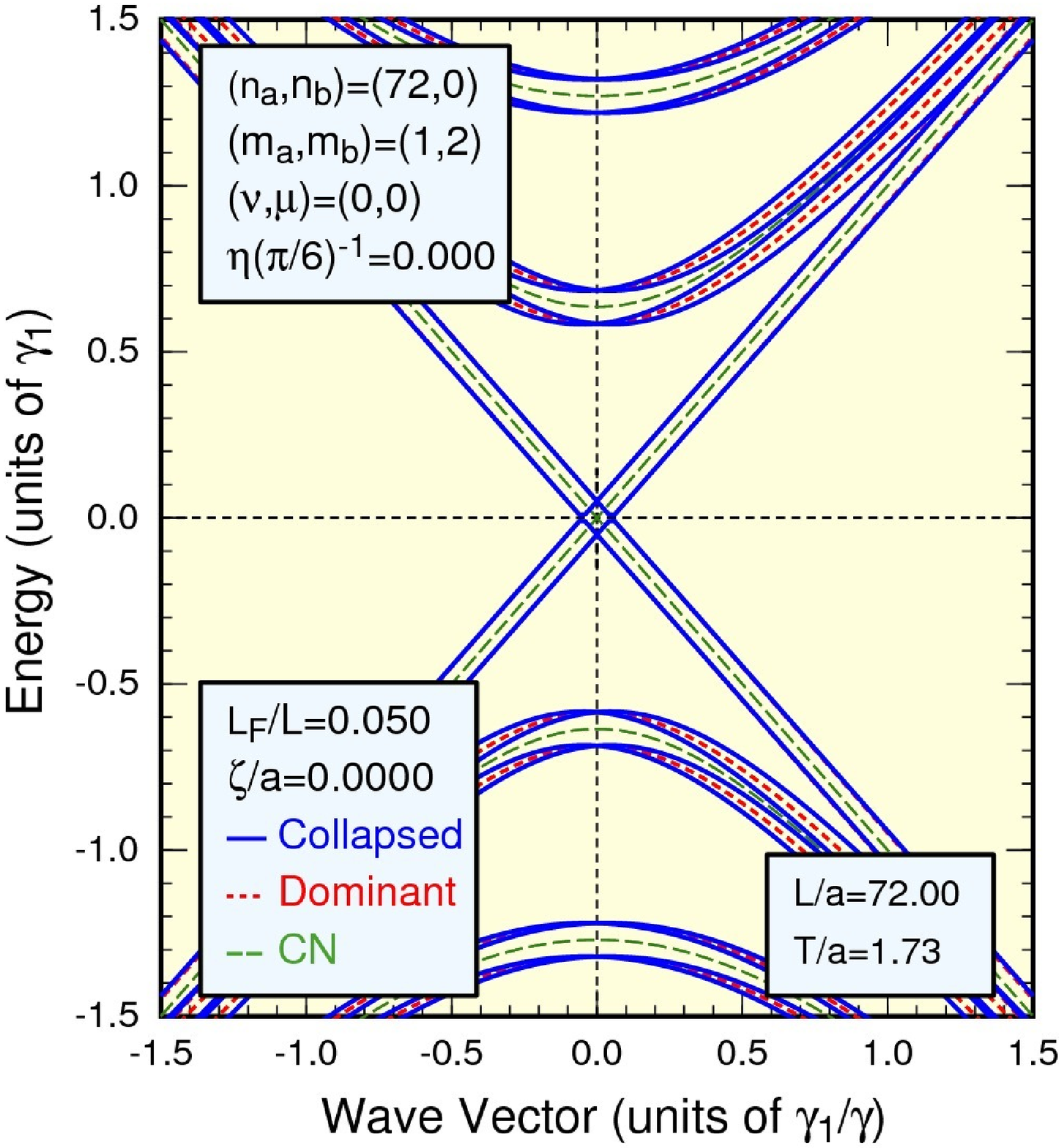}
\enspace ~
(b) \hskip-0.750cm \includegraphics[height=7.50cm]{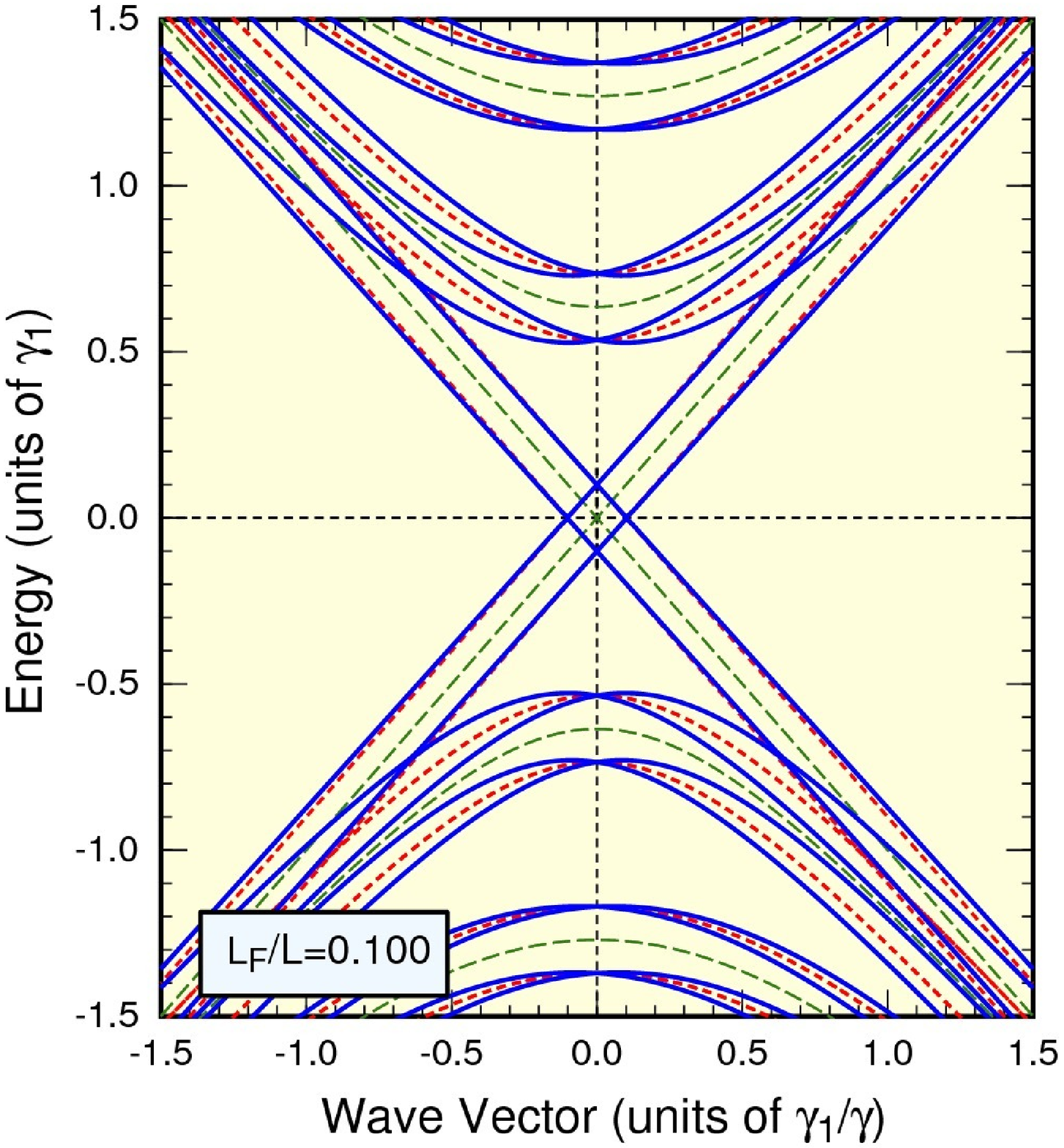}
\end{minipage}
\begin{minipage}[t]{16.0cm}
(c) \hskip-0.750cm \includegraphics[height=7.50cm]{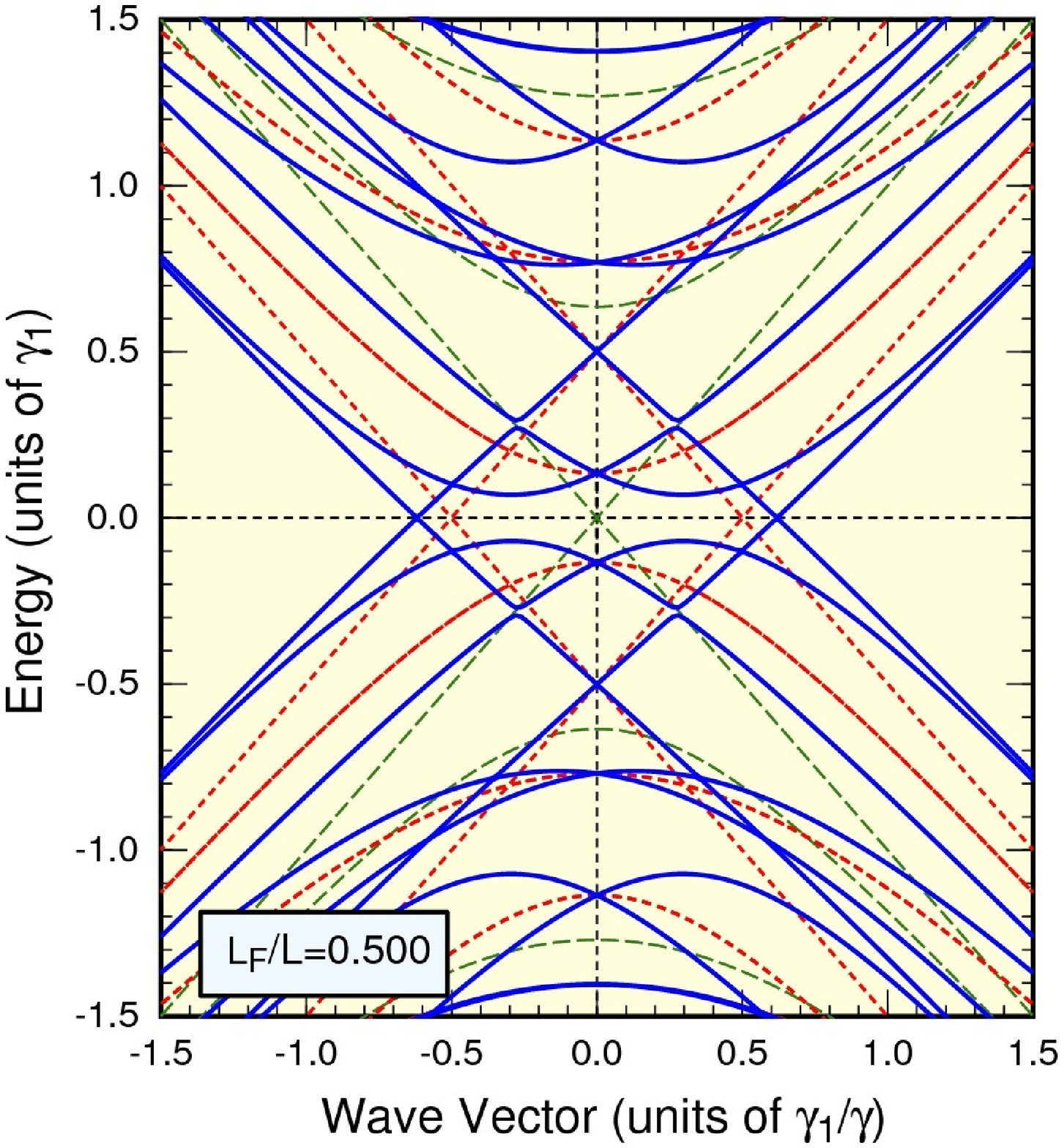}
\enspace
(d) \hskip-0.750cm \includegraphics[height=7.50cm]{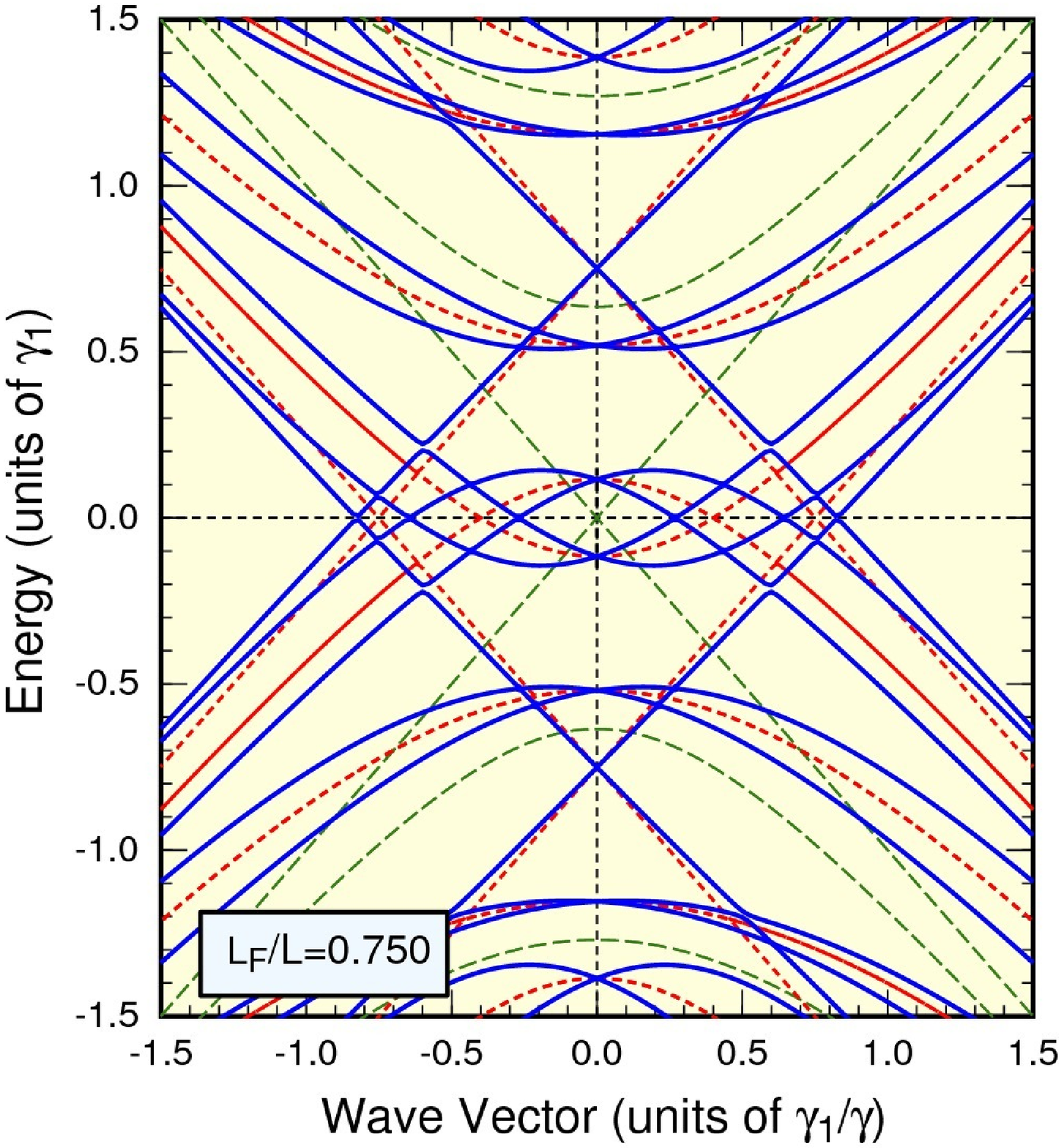}
\end{minipage}
\caption{%
\label{Fig:Band_Structure_for_Varying_Width_(Metallic,_Zigzag,_zeta=0)}
Calculated band structure of collapsed zigzag tubes with $f=144$ and $\eta=0$.
With the increase of $L_F/L$, the band structure gradually approaches that of an AA stacked bilayer with appropriately discretized wave-vectors in the circumference direction.
}
\end{center}
\vspace{-0.50cm}
\end{figure*}
%
\begin{figure*}
\begin{center}
\begin{minipage}[t]{16.0cm}
(a) \hskip-0.750cm \includegraphics[height=7.50cm]{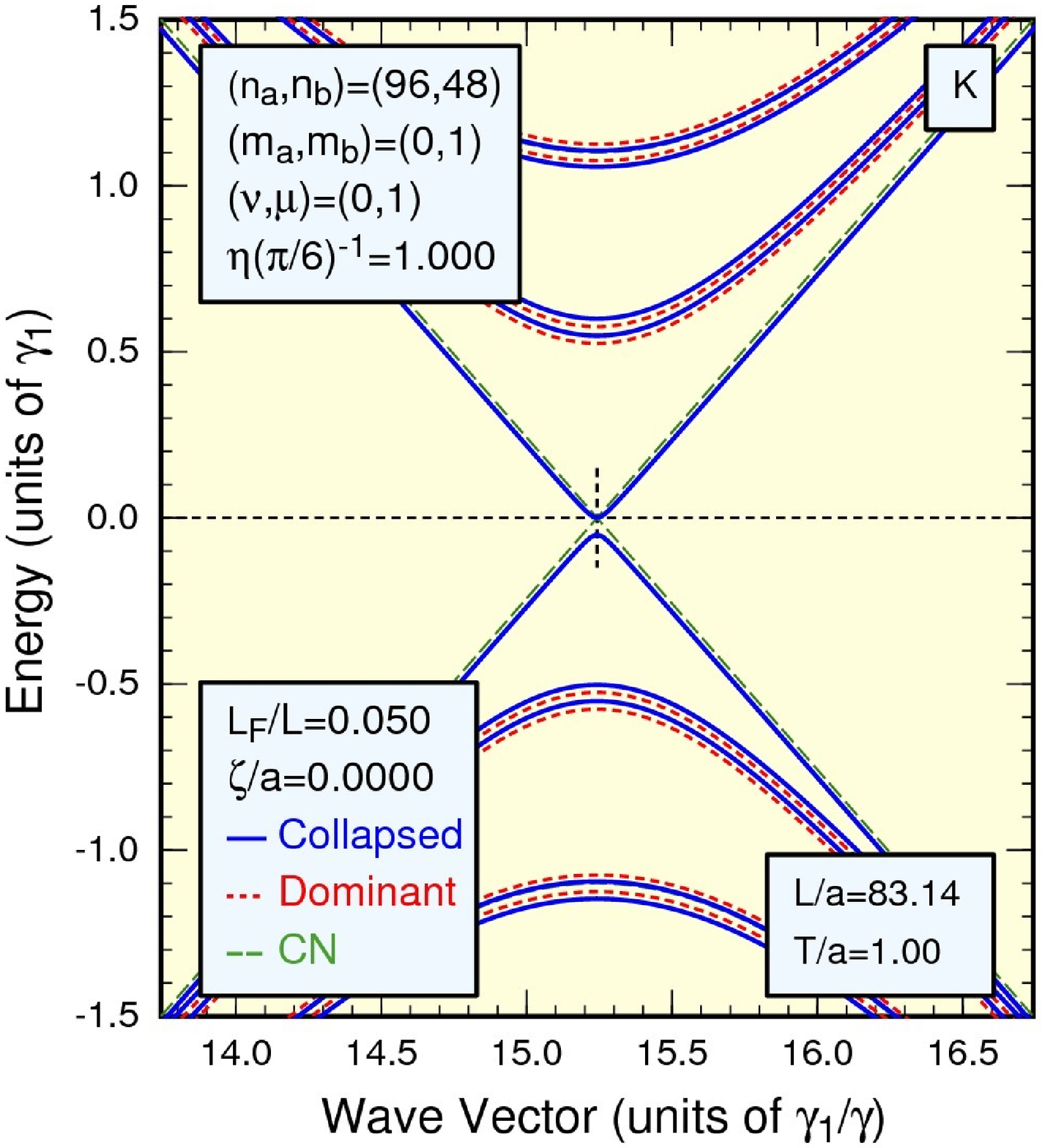}
\enspace ~
(b) \hskip-0.750cm \includegraphics[height=7.50cm]{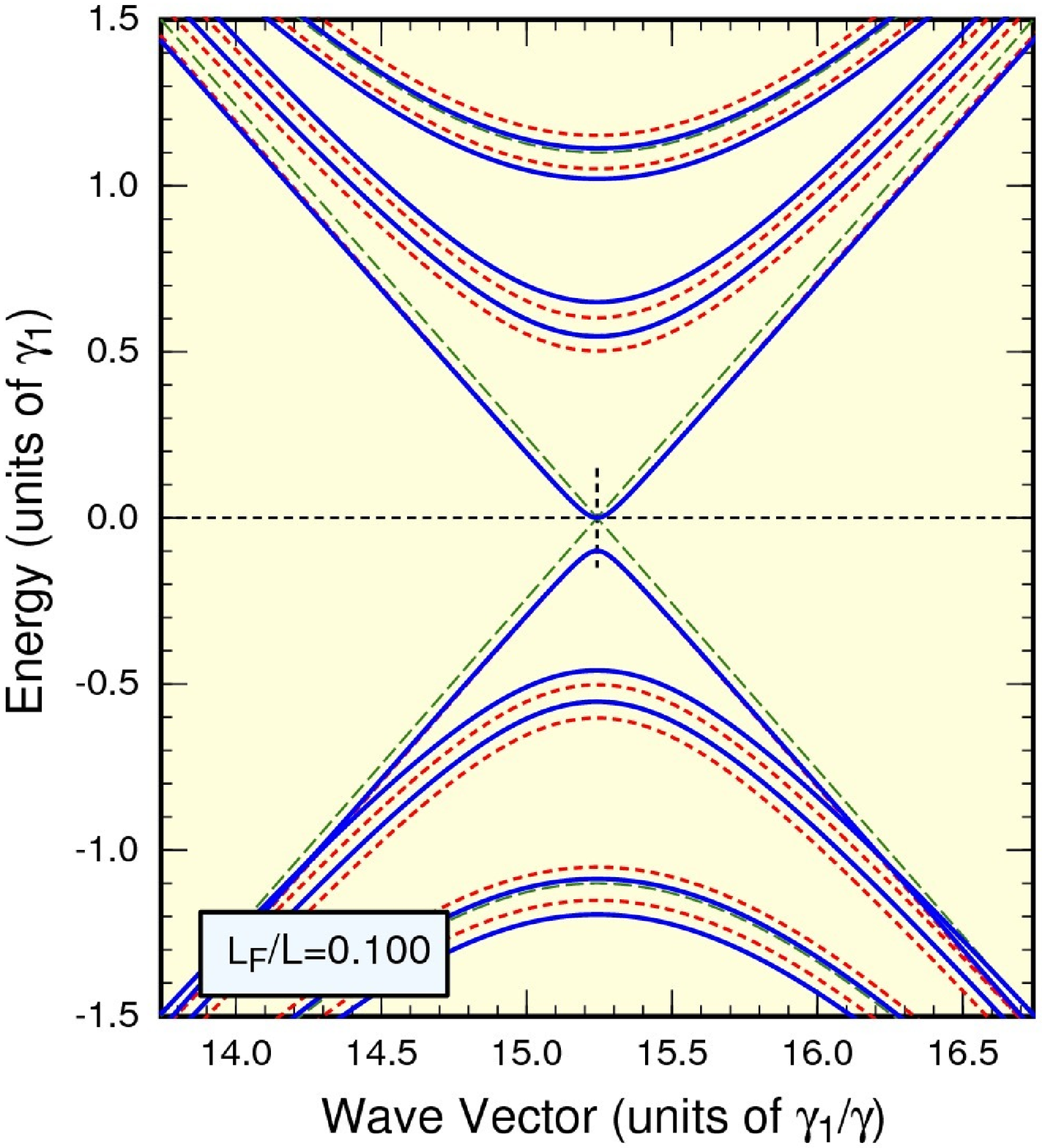}
\end{minipage}
\begin{minipage}[t]{16.0cm}
(c) \hskip-0.750cm \includegraphics[height=7.50cm]{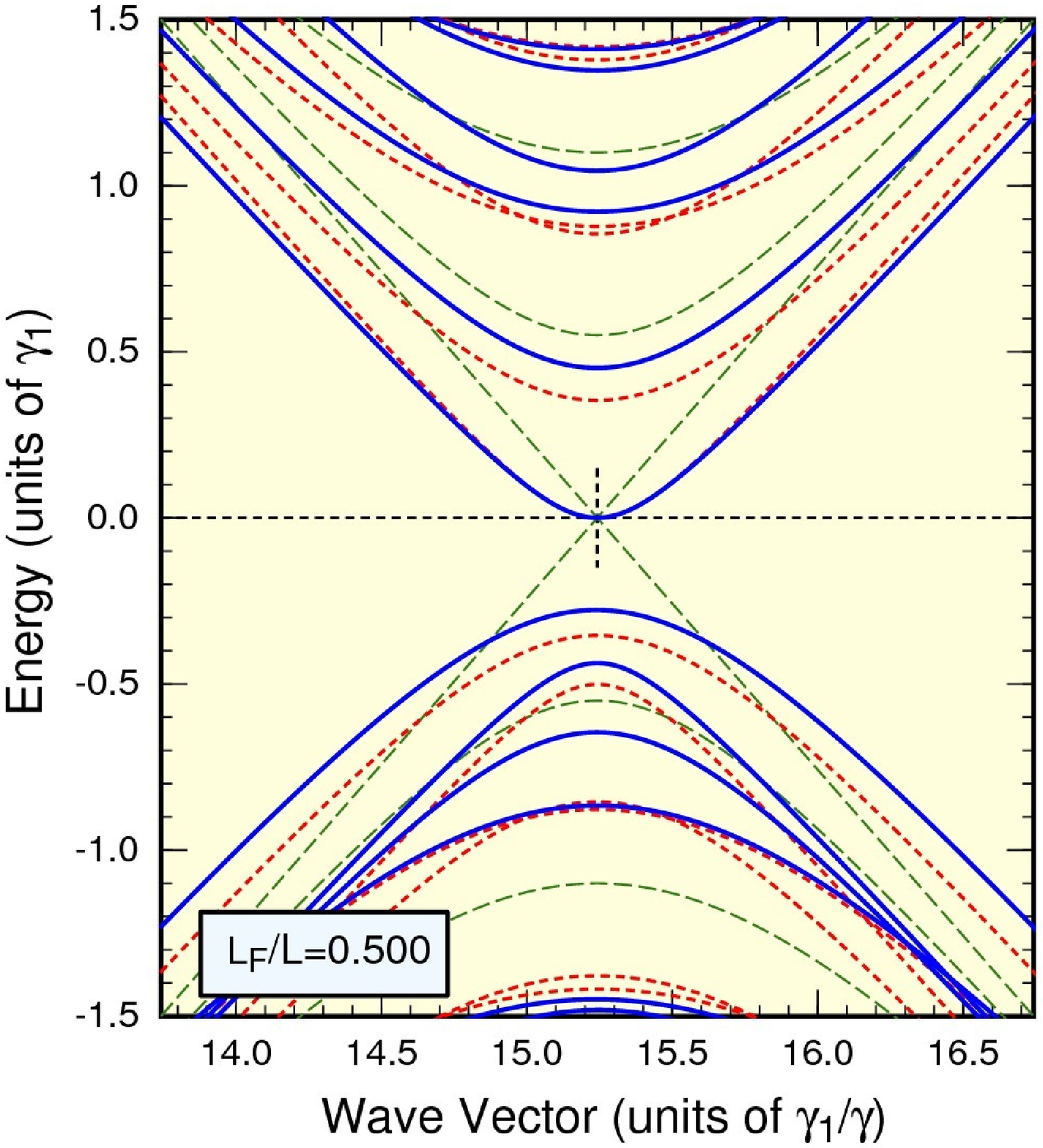}
\enspace
(d) \hskip-0.750cm \includegraphics[height=7.50cm]{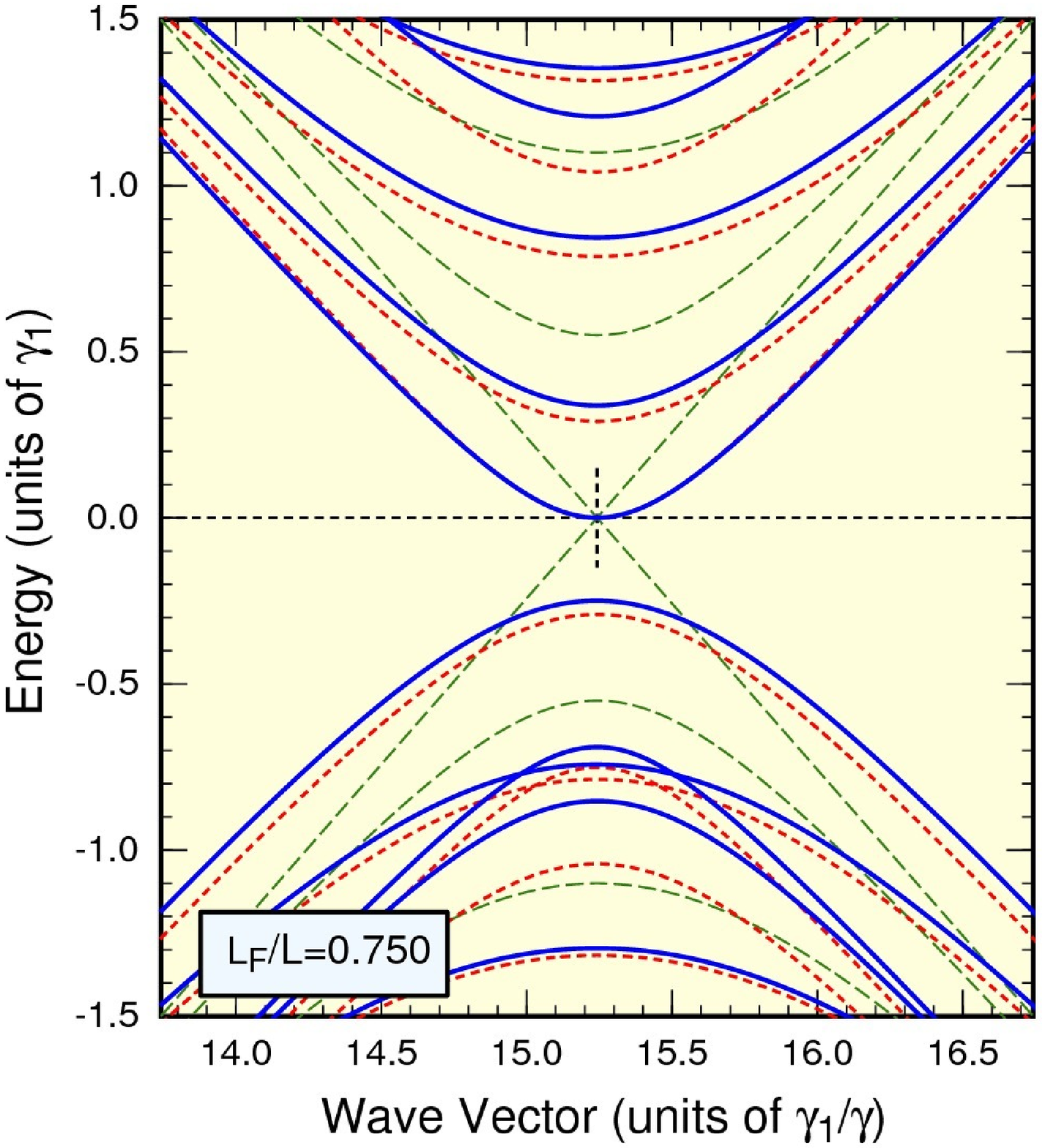}
\end{minipage}
\caption{%
\label{Fig:Band_Structure_for_Varying_Width_(Metallic,_Armchair,_zeta=0)}
Calculated band structure of collapsed armchair tubes with $f=144$ and $\eta=\pi/6$.
With the increase of $L_F/L$, the band structure gradually approaches that of an AB stacked bilayer with appropriately discretized wave-vectors in the circumference direction.
The tube becomes semiconducting due to inter-wall interaction in such a way that the bottom of the conduction band is fixed at zero energy, while the top of the valence band is lowered, forming a band gap.
}
\end{center}
\vspace{-0.50cm}
\end{figure*}
%
\begin{figure*}
\begin{center}
\begin{minipage}[t]{16.0cm}
(a) \hskip-0.750cm \includegraphics[height=7.50cm]{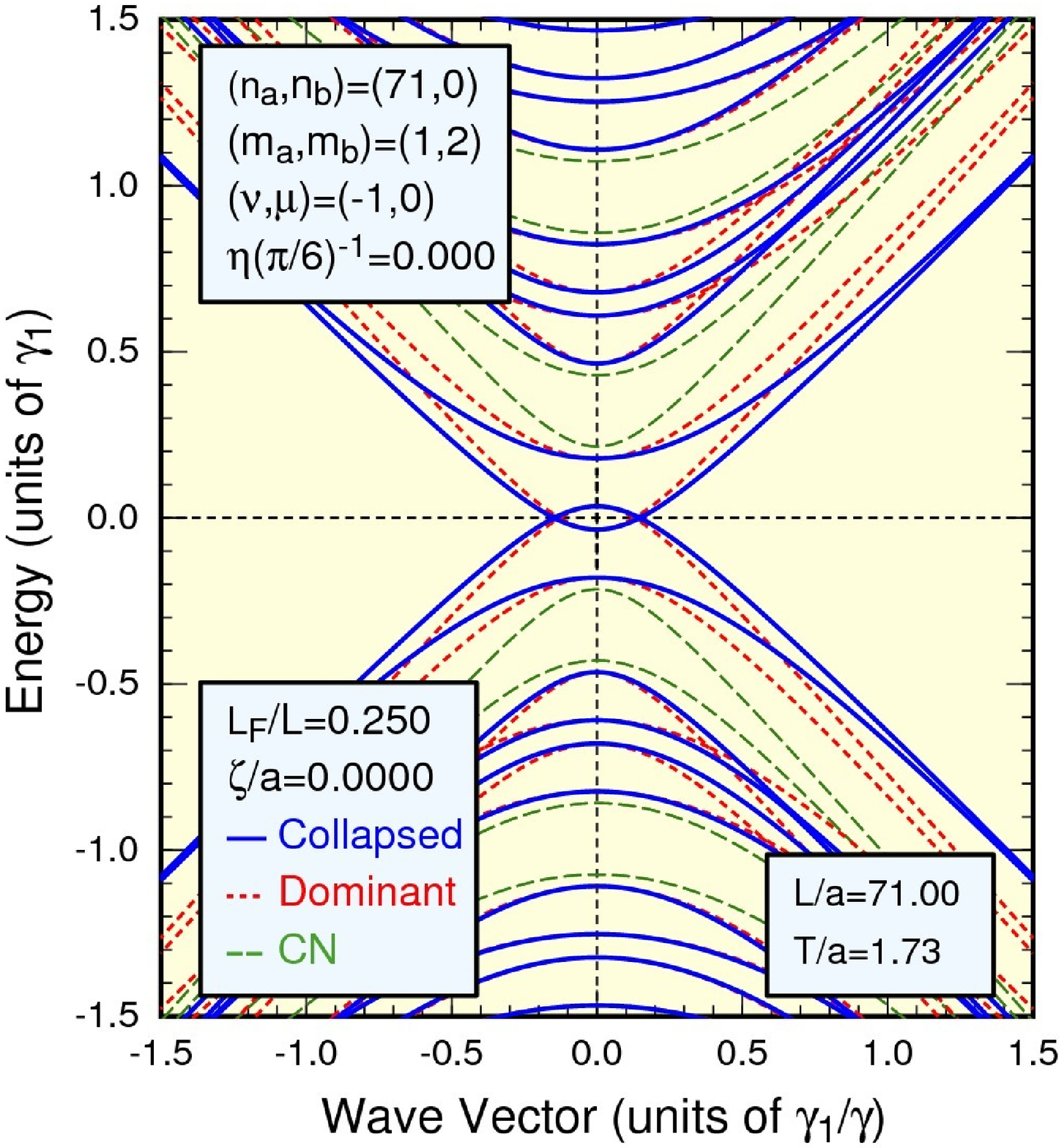}
\enspace ~
(b) \hskip-0.750cm \includegraphics[height=7.50cm]{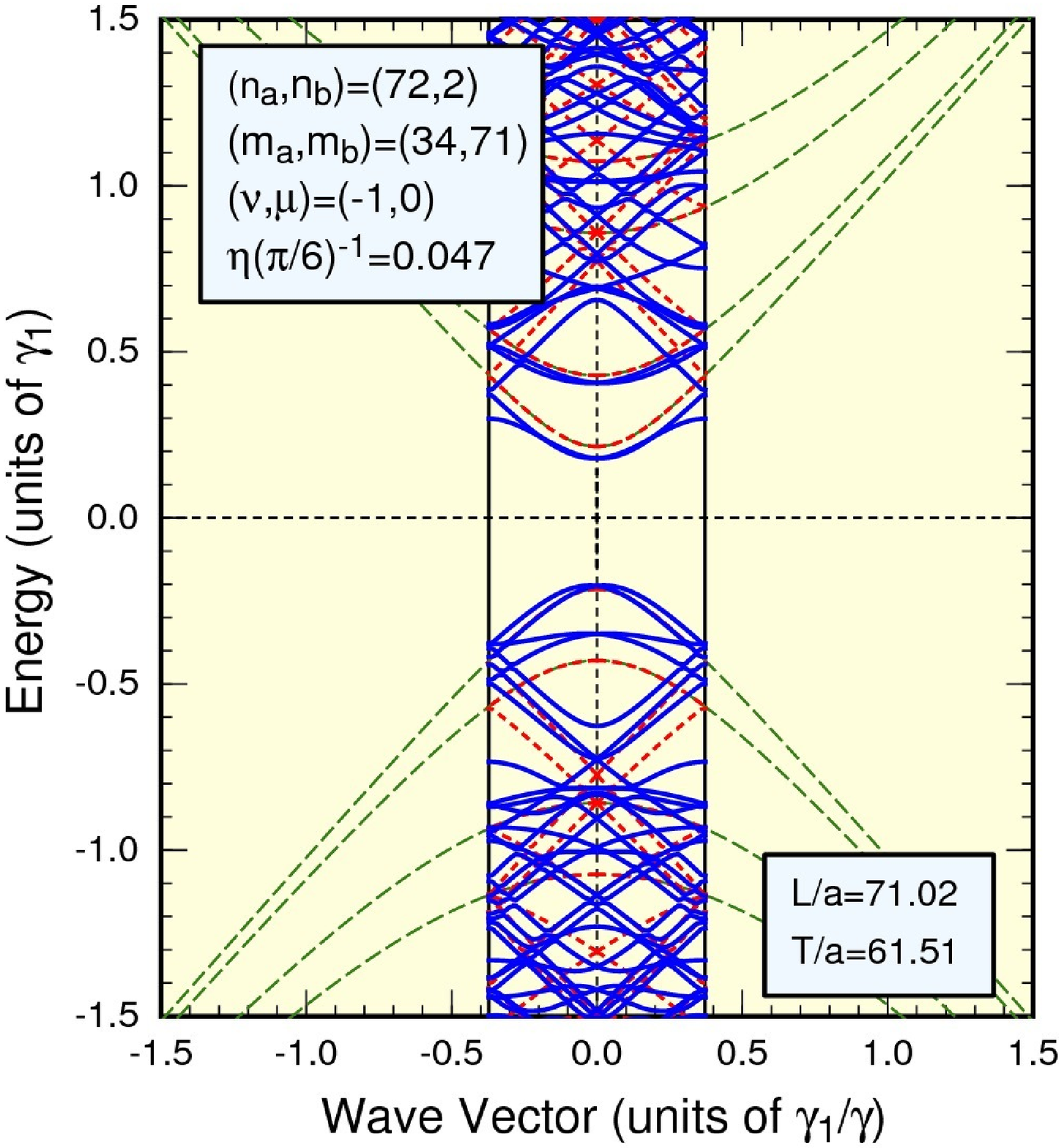}
\end{minipage}
\begin{minipage}[t]{16.0cm}
(c) \hskip-0.750cm \includegraphics[height=7.50cm]{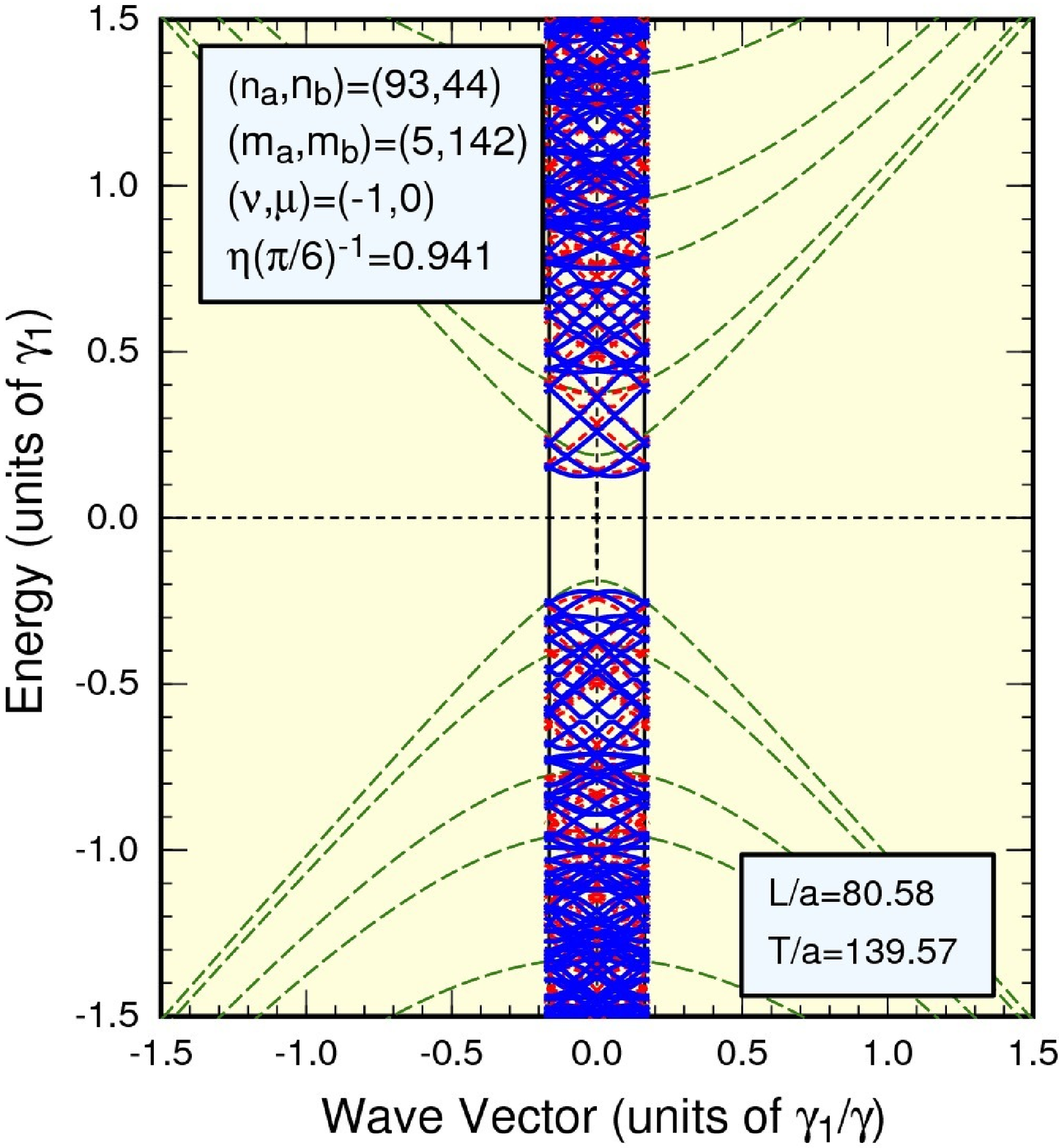}
\enspace
(d) \hskip-0.750cm \includegraphics[height=7.50cm]{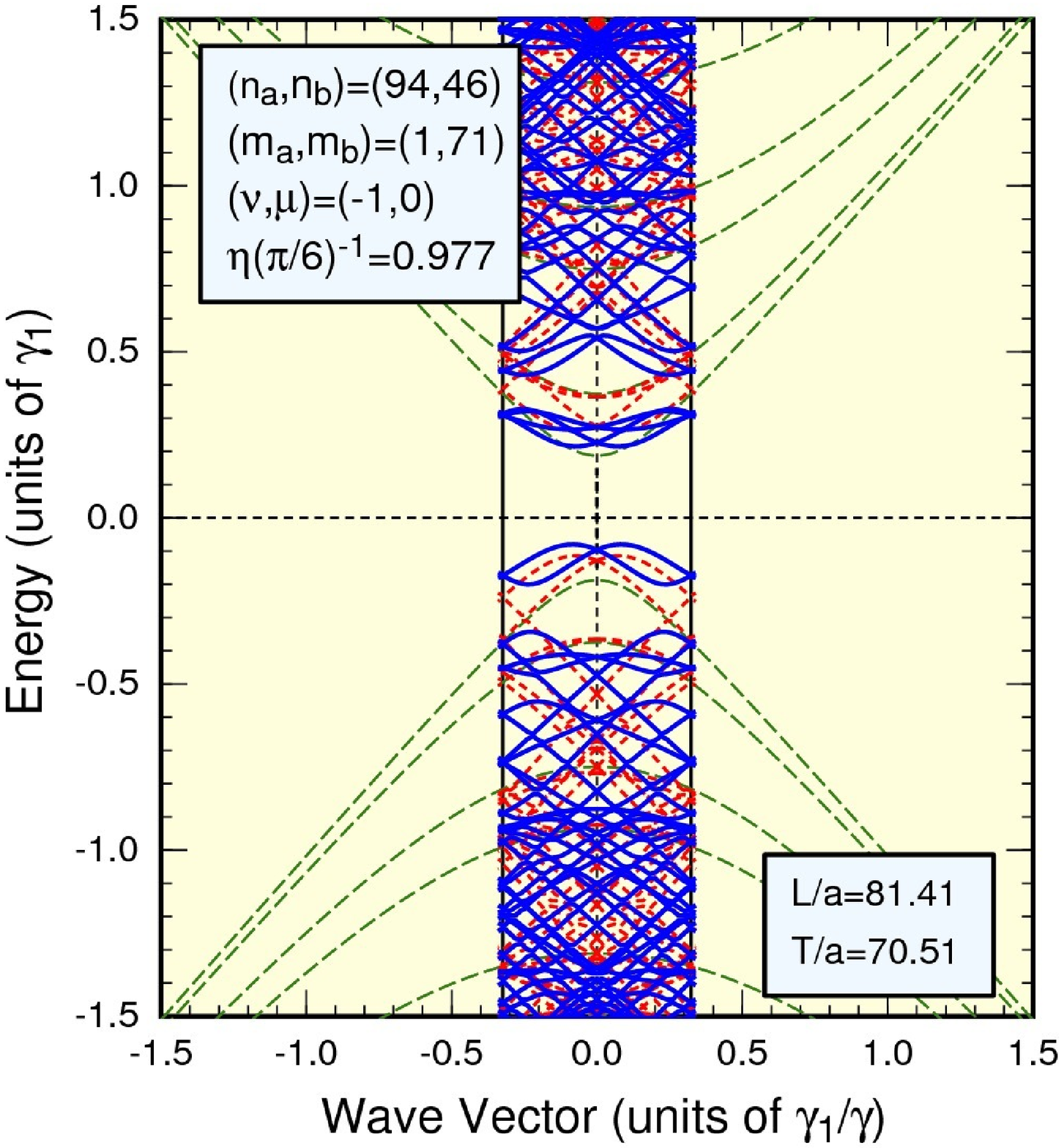}
\end{minipage}
\caption{%
\label{Fig:Band_Structure_(Semiconducting,_Zigzag,_zeta=0)}
Calculated band structure of collapsed tubes with $f=142$ (semiconducting) having (a) zigzag and (b) its neighboring structure, and (c) and (d) near-armchair structure.
The zigzag tube turns into metallic from semiconducting due to collapse as in (a).
}
\end{center}
\vspace{-0.50cm}
\end{figure*}
%
The inter-valley coupling is significant only in the extreme vicinity of a zigzag tube $\eta=0$.
On the other hand, the intra-valley term gradually increases with $\eta$, with behavior strongly dependent on the width of the flattened region, $L_F/L$.
In fact, it oscillates with period roughly proportional to $L/L_F$ ($L$ is a smooth and slowly increasing function of $\eta$), but is not correlated with $T$ that oscillates over wide range as shown in the figures.
Further, we notice that the inter-wall interaction is essentially independent of relative displacement $\zeta$ except at $\eta=0$ (zigzag) and $\pi/6$ (armchair).
This can be seen in the band structure itself as shown in the next section.
\par
%
Calculations are performed also for semiconducting tubes with family number $f=142$, although not shown here.
The behavior is qualitatively the same as in the case of $f=144$, but the absolute value of the effective inter-wall coupling is smaller except in the case of $\eta=0$.
\par
%
In collapsed zigzag tubes with $\eta=0$, inter-wall interaction is present only between the K and K' points.
This corresponds to the fact that phase factors, such as $e^{i({\bf K}'\cdot{\bf R}_A'-{\bf K}\cdot{\bf R}_A)}$ and $e^{i({\bf K}\cdot{\bf R}_A'-{\bf K}'\cdot{\bf R}_A)}$ appearing in the off-diagonal elements of $\tilde V_{AA}$ given in Eq.\ (\ref{Eq:VAA}) and the corresponding terms in $\tilde V_{AB}$, etc.\ in Eqs.\ (\ref{Eq:VAB})--(\ref{Eq:VBB}), do not cancel out even after summation over ${\bf R}_A$, ${\bf R}_A'$, etc.
When $\eta$ slightly deviates from zero, however, these phase factors start to rapidly oscillate in a quasi-periodic manner because they involve ${\bf K}$ and ${\bf K}'$ considerably different ($\sim2\pi/a)$ from each other.
Thus, the effective inter-wall potential vanishes due to cancellation.
\par
%
In collapsed armchair tubes with $\eta=\pi/6$, on the other hand, the relevant phase factors involve same ${\bf K}$ or ${\bf K}'$.
When $\eta$ slightly deviates from $\pi/6$, the phase factors start to oscillate in a quasi-periodic manner but the oscillation is relatively slowly-varying.
Thus, the effective potential remains nonzero for small $L_F$ because of incomplete cancellation.
Further, it decreases with the deviation of $\eta$ from $\pi/6$ more rapidly for wider flattened region, as shown in Fig.\ \ref{Fig:Dominant-Term_Approximation:_1}.
\par
%
\section{Numerical Results} \label{Sec:Numerical_Results}
%
For actual calculations, we choose $\gamma_1$ as the energy unit.
Further, we choose $n=0,\pm1,\cdots,\pm n_{\rm max}$ and for $m=0,\pm1,\cdots,\pm m_{\rm max}$, with $n_{\rm max}=m_{\rm max}=10\sim15$.
This choice of the basis set gives convergent results at least for the bands lying in the zero-energy region in which we are interested.
In the following, results for $\zeta/a=0$, 1/4, and $1/\sqrt3$ will be shown.
Calculated band structure will be compared with that in the dominant-term approximation in which only the dominant term is taken as discussed in the previous section.
\par
%
Figure \ref{Fig:Band_Structure_(Metallic_zeta=0)} shows examples of the band structure in the metallic case with $f=144$ for $\zeta=0$.
Figures \ref{Fig:Band_Structure_(Metallic_zeta=0)} (a) and (b) present those in the vicinity of the zigzag structure, i.e., $(n_a,n_b)=(72,0)$ and (73,2) corresponding to $\eta(\pi/6)^{-1}=0$ and 0.046, respectively.
The dominant-term approximation can describe the essential features of the bands near the Fermi level consisting of metallic linear bands, although wave vectors corresponding to zero energy are shifted and the velocity is lowered if we go beyond the dominant-term approximation.
\par
%
In zigzag nanotubes, the band structure is strongly modified by collapsing due to the strong inter-wall couplings, although the tube remains metallic because of the presence of linear bands at the Fermi level.
In fact, the metallic linear bands associated with the K and K' points are split in energy or shifted in the positive and negative $k$ direction due to the inter-wall coupling as shown in Eq.\ (\ref{Eq:Perturbation_Zigzag_Metallic_Linear}) in the perturbation treatment in the previous section.
The excited parabolic bands which are four-fold degenerate are first split into two sets at $k=0$ and then the remaining degeneracy is lifted by the $k$ linear term as shown in Eq.\ (\ref{Eq:Perturbation_Zigzag_Metallic_Parabolic}).
\par
%
With the increase of $\eta$, i.e., when the structure deviates from the zigzag case, effects of inter-wall interaction rapidly diminish.
In fact, for $(n_a,n_b)=(73,2)$, shown in Fig.\ \ref{Fig:Band_Structure_(Metallic_zeta=0)} (b), the band structure is modified due to the collapse in such a way that the effective velocity in the axis direction is slightly reduced.
This velocity reduction is in qualitative agreement with that observed experimentally\cite{de_Heer_et_al_2007a,Hass_et_al_2008a,Sprinkle_et_al_2009a,Ni_et_al_2008c,Schmidt_et_al_2010a,de_Heer_et_al_2010a,de_Heer_et_al_2011a,Luican_et_al_2011a,Brihuega_et_al_2012a} and calculated theoretically\cite{Lopes_dos_Santos_et_al_2007a,Latil_et_al_2007a,Hass_et_al_2008a,Shallcross_et_al_2008a,Bistritzer_and_MacDonald_2011a,Bistritzer_and_MacDonald_2011b,Lopes_dos_Santos_et_al_2012a} in twisted bilayer graphene.
With the increase of $\eta$, however, the band rapidly becomes unaffected by collapsing, although the results are not shown here.
For chiral tubes, $\mu$ takes a nonzero value and therefore the K and K' points become different in the one-dimensional Brillouin zone as denoted by short vertical dotted lines near zero energy.
\par
%
Figure \ref{Fig:Band_Structure_(Metallic_zeta=0)} also shows results for tubes having a structure close to $\eta=\pi/6$ (armchair) and for $\zeta=0$, i.e., (c) $(n_a,n_b)=(94,42)$, (d) (96,44), (e) (95,46), and (f) (96,48) corresponding to (c) $\eta(\pi/6)^{-1}=0.893$, (d) 0.930, (e) 0.965, and (f) 1.
In the armchair tube, the figure shows results only for the K point and those for the K' point are obtained by mirror reflection with respect to $k=0$.
\par
%
In an armchair tube shown in Fig.\ \ref{Fig:Band_Structure_(Metallic_zeta=0)} (f), the metallic band structure is strongly modified by collapsing and the tube becomes semiconducting due to band-gap opening.
In fact, the bottom of the conduction band with $n=0$ remains at zero energy, while the top of the valence band with $n=0$ is lowered roughly in proportion to $L_F/L$ due to inter-wall coupling.
This is in qualitative agreement with Eq.\ (\ref{Eq:Perturbation_Armchair_AB_Linear}) obtained by the perturbation analysis.
For excited parabolic bands, the qualitative features of effects of inter-wall coupling are in agreement with the perturbation analysis giving Eq.\ (\ref{Eq:Perturbation_Armchair_AB_Parabolic}).
\par
%
In chiral nanotubes effects of inter-wall interactions are considerably reduced and diminish with the decrease of $\eta$ from $\pi/6$, although their decay is more gradual than in the vicinity of the zigzag tube.
This has already been demonstrated in the behavior of the dominant terms shown in Fig.\ \ref{Fig:Dominant-Term_Approximation:_1}.
The dominant-term approximation gives quite accurate results near zero energy, but starts to become less valid away from zero energy.
\par
%
The corresponding results for $\zeta/a=1/4$ and $1/\sqrt3$ are shown in Fig.\ \ref{Fig:Band_Structure_(Metallic,_Zigzag,_zeta=1/4,_1/sqrt3)}.
Because the band structure is not affected by displacement $\zeta$ in chiral nanotubes, only the results for zigzag and armchair nanotubes are shown.
In zigzag and armchair nanotubes, the band structure depends significantly on $\zeta/a$.
\par
%
As shown in Fig.\ \ref{Fig:Structure_of_Flattened_Zigzag_Nanotube}, the zigzag tube with $\eta=0$ has the structure of an AA stacked bilayer graphene in the flattened region for $\zeta/a=0$ and varies as a function of $\zeta$ with period $a/2$.
For $\zeta/a=1/4$, the two layers are displaced from each other in a symmetric way, resulting in the symmetric band structure as shown in Fig.\ \ref{Fig:Band_Structure_(Metallic,_Zigzag,_zeta=1/4,_1/sqrt3)} (a).
A small band gap appears for $L_F/L=1/4$, but disappears for sufficiently large $L_F/L$, although explicit results are not shown here.
\par
%
The displacement $\zeta/a=1/\sqrt3=0.5774\cdots$ in the zigzag case, shown in Fig.\ \ref{Fig:Band_Structure_(Metallic,_Zigzag,_zeta=1/4,_1/sqrt3)} (c), corresponds to the case that the top and bottom layers are slightly displaced from an AA stacked bilayer.
This slight displacement results in repulsion between some bands of Fig.\ \ref{Fig:Band_Structure_(Metallic_zeta=0)} (a), giving rise to the band-gap opening.
This gap due to symmetry breaking is always present independent of $L_F/L$.
\par
%
As shown in Fig.\ \ref{Fig:Structure_of_Flattened_Armchair_Nanotube}, for the armchair nanotube, the structure takes the form of AB stacking at $\zeta/a=0$ and $\zeta/a=1/(2\sqrt3)=0.2886\cdots$, and then the form of AA stacking at $\zeta/a=1/\sqrt3$.
This change repeats itself with period $a/\sqrt3$.
Thus, in an armchair tube with $\zeta/a=0.25$, the structure is slightly displaced from the AB stacking.
As shown in Fig.\ \ref{Fig:Band_Structure_(Metallic,_Zigzag,_zeta=1/4,_1/sqrt3)} (b), this slight displacement results in some distortion of the band structure of Fig.\ \ref{Fig:Band_Structure_(Metallic_zeta=0)} (f) in such a way that the energy becomes asymmetric around the K point.
In spite of the asymmetry, the tube remains semiconducting due to nonzero gap.
\par
%
For $\zeta/a=1/\sqrt3$, the flattened region has the structure of an AA stacked bilayer, and nanotubes become metallic independent of the width of the flattened region, as shown in Fig.\ \ref{Fig:Band_Structure_(Metallic,_Zigzag,_zeta=1/4,_1/sqrt3)} (d), because linear bands cross the Fermi level.
In agreement with Eq.\ (\ref{Eq:Perturbation_Armchair_AA_Linear}), the metallic linear bands are shifted in the negative $k$ direction.
The parabolic bands are split and shifted in different $k$ directions depending on band $n$ qualitatively in agreement with Eq.\ (\ref{Eq:Perturbation_Armchair_AA_Parabolic}).
\par
%
In Fig.\ \ref{Fig:Band_Structure_for_Varying_Width_(Metallic,_Zigzag,_zeta=0)}, the dependence on $L_F/L$ is shown for zigzag tubes with $\zeta/a=0$.
With the increase of $L_F/L$, the spectrum gradually takes a form of that of an AA stacked bilayer with appropriately discretized wave-vectors perpendicular to the axis.
The tube remains metallic independent of $L_F/L$.
The dependence on the width of the flattened region for armchair nanotubes is shown in Fig.\ \ref{Fig:Band_Structure_for_Varying_Width_(Metallic,_Armchair,_zeta=0)} for $\zeta/a=0$.
The band structure again gradually approaches that of an AB stacked bilayer.
The band gap increases, takes a maximum, and then decreases with $L_F/L$, but always remains nonzero.
\par
%
As some examples for semiconducting nanotubes, we shall consider the case of $f=142$ corresponding to $(n_a,n_b)=(71,0)$, $(72,2)$, $\dots$, $(94,46)$.
In this case we always have $\nu=-1$ and $\mu=0$, i.e., the K and K' points are both mapped onto the center of the one-dimensional Brillouin zone.
Because qualitative feature of the dependence on the chiral angle is the same as in metallic nanotubes, we shall present results in the vicinity of the zigzag and armchair structure in Fig.\ \ref{Fig:Band_Structure_(Semiconducting,_Zigzag,_zeta=0)}.
In fact, inter-wall interactions rapidly become small with the increase of $\eta$ from $\eta=0$.
The same is true for $\eta\!\sim\!\pi/6$, i.e., inter-wall effects are most important for $(n_a,n_b)=(94,46)$ for which $\eta=0.977\!\times\!(\pi/6)$ closest to the armchair structure and decrease with the decrease of $\eta$ although more slowly.
\par
%
One most significant effect of the collapse is to convert semiconducting into metallic in the zigzag tube.
This arises due to the splitting of two bands degenerate between the K and K' points due to inter-wall coupling as has been shown in the perturbation analysis, Eq.\ (\ref{Eq:Perturbation_Zigzag_Semiconducting}).
All chiral tubes remain semiconducting independent of $L_F/L$, although explicit results are not shown.
\par
%
\section{Summary and Conclusion} \label{Sec:Summary_and_Conclusion}
%
We have theoretically studied effects of inter-wall interaction in collapsed carbon nanotubes within an effective-mass scheme.
Inter-wall interactions in the flattened region are represented by an effective potential connecting a point on the flattened region to its counter point.
Effects of inter-wall interactions are most important in nonchiral nanotubes such as zigzag and armchair.
In zigzag and armchair tubes, the band structure varies sensitively with the displacement, corresponding to the sensitive change of the band structure in bilayer graphene.
In zigzag nanotubes, in particular, the collapsed tubes become metallic for sufficiently wide flattened region independent of whether the uncollapsed tube is metallic or semiconducting.
\par
%
In chiral nanotubes, inter-wall interactions can essentially be neglected except in the close vicinity of zigzag and armchair tubes.
Inter-wall interactions diminish rapidly when chiral angle deviates from 0 (zigzag) or $\pi/6$ (armchair), although the decay is slower in the vicinity of the armchair tube.
In fact, in chiral tubes closest to a zigzag and armchair tube, the semiconducting tube remains semiconducting even for very wide flattened region.
Such qualitative features of the chiral angle dependence can be understood through the magnitude of dominant terms corresponding to long-wavelength Fourier coefficients of the effective inter-wall potential.
\par
%
Small band gap is inversely proportional to diameter in thick chiral semiconducting nanotubes, and is smaller than $\gamma_1=0.4$ eV in nonchiral nanotubes, when band gap opens due to inter-layer interaction.
Observation of these band gaps is required by means of precise measurement such as infrared transmission spectroscopy and scanning tunneling microscopy.
Slightly reduced velocity due to the inter-layer interaction may be observed with scanning tunneling microscopy, angle-resolved photoemission spectroscopy, and Raman spectroscopy in the same way as in twisted bilayer graphene.\cite{de_Heer_et_al_2007a,Hass_et_al_2008a,Sprinkle_et_al_2009a,Ni_et_al_2008c,Schmidt_et_al_2010a,de_Heer_et_al_2010a,de_Heer_et_al_2011a,Luican_et_al_2011a,Brihuega_et_al_2012a}
\par
%
With the increase in the width of the flattened region, the band structure approaches that of a bilayer ribbon in which the electron motion in the ribbon-width direction is discretized under appropriate boundary conditions.
Therefore, the band structure of collapsed nanotubes can be obtained from a bilayer graphene by introducing appropriate boundary conditions corresponding to the curved monolayer region.
This problem is left for future study.
\par
%
%
\begin{center}\vbox{\vspace{0.50cm}{\bf Acknowledgments}\vspace{0.0cm}}\end{center}
%
This work has been supported in part by MEXT Grants-in-Aid for Scientific Research on Innovative Areas ``Science of Atomic Layers'' (Project No.\ 2506, 26107534) and Scientific Research (Project No.\ 24540339) in Japan.
\par
%

%
\bigskip
\hrule
\smallskip
\rightline{\jobname.tex (\today)}
%
\end{document}